\documentclass[aps,prd,reprint,groupedaddress]{revtex4-1}
\pdfoutput=1

\usepackage{amsmath}
\usepackage[utf8]{inputenc}
\usepackage{multirow}
\usepackage{graphicx}
\usepackage{latexsym}
\usepackage[colorlinks]{hyperref}
\hypersetup{allcolors=[rgb]{0,0,1}}
\usepackage[dvipsnames]{xcolor} 
\usepackage{xfrac}

\graphicspath{{Final_Fig/}}

\begin{document}

\title{Effects of SASI and LESA in the neutrino emission of rotating supernovae}

\author{Laurie Walk}
\affiliation{Niels Bohr International Academy and DARK, Niels Bohr Institute, University of Copenhagen, Blegdamsvej 17, 2100, Copenhagen, Denmark}
\author{Irene Tamborra}
\affiliation{Niels Bohr International Academy and DARK, Niels Bohr Institute, University of Copenhagen, Blegdamsvej 17, 2100, Copenhagen, Denmark}
\author{Hans-Thomas Janka}
\affiliation{Max-Planck-Institut f\"ur Astrophysik,
 Karl-Schwarzschild-Str.~1, 85748 Garching, Germany}
 \author{Alexander Summa}
\affiliation{Max-Planck-Institut f\"ur Astrophysik,
 Karl-Schwarzschild-Str.~1, 85748 Garching, Germany}

\date{\today}
\begin{abstract}
Rotation of core-collapse supernovae (SNe) affects the neutrino emission characteristics. By comparing the neutrino properties of three  three-dimensional SN simulations of a $15\,M_\odot$ progenitor (one non-rotating model and two models rotating at different velocities), we investigate how the neutrino emission varies with the flow dynamics in the SN core depending on the different degrees of rotation. The large-amplitude sinusoidal modulations due to the standing accretion-shock instability (SASI) are weaker in both the rotating models than in the non-rotating case. The SN progenitor rotation  reduces the radial velocities and radial component of the kinetic energy associated with convection interior to the  proto-neutron star. This effect seems to  disfavor the growth of the hemispheric neutrino-emission asymmetries associated with the lepton-emission self-sustained  asymmetry (LESA). An investigation of the multipole expansion of the neutrino luminosity and the electron neutrino lepton number flux shows a  dominant quadrupolar mode in rotating SN models. Our findings highlight the power of using neutrinos as probes of  SN hydrodynamics.
\end{abstract}

\maketitle
\section{Introduction} 

Neutrinos are key particles in the yet mysterious  core-collapse supernova (SN) explosion mechanism, carrying about $99\%$ of the total explosion energy~\cite{Janka:2016fox}. According to the delayed neutrino mechanism, neutrinos revive the shock-wave stalled in the iron shell which finally triggers the explosion. Hydrodynamical instabilities take place in this phase, enhancing the efficiency of the energy transferred to the shock-wave for its revival. In particular,  large-scale convective overturns develop in the neutrino-heated post-shock layer as well as the standing accretion shock instability (SASI), involving a global deformation and sloshing motion of the shock front. 
 
The advent of high-fidelity SN modelling in three dimensions (3D) has recently begun~\cite{Hanke2013,Tamborra:2014hga,Janka:2016fox,Summa:2017wxq,Melson2015a,Melson:2015spa,Mueller2015a,Muller:2016izw,Chan:2017tdg,Takiwaki2014a,Takiwaki:2016qgc,Muller:2017hht,Lentz:2015nxa,Ott:2017kxl}; it has shed new light on the strong directional dependence of the neutrino emission properties~\cite{Tamborra:2014hga,Walk:2018gaw,Tamborra:2014aua,Tamborra:2013laa,Lund:2012vm,Takiwaki:2017tpe,Kuroda:2017trn}.
Notably, a dedicated post-processing of the neutrino emission characteristics from 3D SN simulations,  performed with sophisticated neutrino transport,  illustrated a significant sensitivity of the neutrino signal to the SASI instability~\cite{Tamborra:2014hga,Tamborra:2014aua,Tamborra:2013laa,Lund:2012vm,Lund:2010kh}. It was found that the neutrino signal from high-mass progenitors is likely to exhibit imprints of  SASI~\cite{Tamborra:2013laa,Tamborra:2014hga} in the form of  large-amplitude sinusoidal modulations detectable for observers located in the proximity of the SASI plane, even for SNe at the edge of our Galaxy. In particular, the Fourier power spectrum of the detectable neutrino signal was found to exhibit a clear peak  corresponding to the SASI oscillation frequency~\cite{Tamborra:2013laa}. On the other hand, low mass progenitors seem to be characterized by pure convective activity which  induces small-scale and low amplitude modulations in the neutrino signal not detectable experimentally~\cite{Tamborra:2013laa,Tamborra:2014hga}. 

Three dimensional simulations with energy-dependent neutrino-transport also led to the discovery of the first hydrodynamical instability fostered  by neutrinos, the so-called lepton-emission self-sustained asymmetry (LESA)~\cite{Tamborra:2014aua}. This surprising phenomenon manifests itself as a macroscopic dipole in the electron neutrino lepton number (ELN) which may have interesting consequences on neutrino flavor conversions, the nucleosynthesis of chemical elements, and the neutron-star radius.  LESA has recently been confirmed by large and independent sets of simulations~\cite{Janka:2016fox,Glas:2018vcs,Vartanyan:2018iah,OConnor:2018tuw,Powell:2018isq}. 

Another crucial aspect, potentially affecting the (detectable) neutrino signal, is the progenitor rotation. The role of SN angular momentum is a subject of debate. In particular, rotation has been invoked as a crucial ingredient  favoring the SN explosion, potentially affecting the production of $r$-process elements, and determining the pulsar spin period. Rotation is also considered as relevant for the so-called magneto-rotational mechanism, according to which magnetic fields are amplified to dynamically relevant strengths by extremely high spin rates \cite{Mosta:2014jaa,Nishimura2015,Obergaulinger2017,Obergaulinger2018}. 
The progenitor rotation also seems to affect the neutrino emission properties~\cite{Walk:2018gaw,Takiwaki:2017tpe} as well as LESA, which is expected to be damped by progenitor rotation~\cite{Janka:2016fox}.

By relying on the three self-consistent 3D hydrodynamical simulations of a $15\,M_\odot$ SN (a non-rotating model, and two models rotating at different velocities within observational constraints) of Ref.~\cite{Summa:2017wxq}, we investigated  the possibility of adopting neutrinos as SN gyroscopes in a companion paper~\cite{Walk:2018gaw}. It was concluded that the detectable  modulations in the neutrino event rate due to SASI are smeared out by rotation. In particular, fast rotation led to a successful explosion, and long-lasting accretion downflows  of matter affected the detectable neutrino signal after the explosion. 

By looking at the spectrograms of the detectable neutrino event rate for the three SN models,  it was found that rotation is responsible for inducing modulations in the neutrino rate at frequencies higher than the SASI frequency~\cite{Walk:2018gaw}. Such a feature is a discriminant of the progenitor rotation and it should be detectable for observers located in the proximity of the main SASI plane. On the other hand, if the observer is located away from the SASI plane, degeneracies in the neutrino event rate due to weak SASI modulations and the effects of the progenitor rotation may occur, making it more difficult to determine the progenitor rotational velocity through neutrinos. 

In this work, we analyze the phenomenological implications of  SN rotation on the neutrino emission characteristics, in particular connected to SASI
and LESA features. We stress that, given the complex interplay of the various hydrodynamical instabilities, our work is a first step towards  pinpointing the effects of rotation through neutrinos; an in-depth analysis of the consequences of rotation on the hydrodynamic instabilities in SN cores is beyond the scope of the present work. Our study is based on the non-rotating and two  rotating $15\,M_\odot$ SN models presented in Ref.~\cite{Summa:2017wxq}. The new results are compared with those obtained for a set of progenitors and their SN simulations presented in 
previous publications~\cite{Hanke2013,Tamborra:2013laa,Tamborra:2014hga,Tamborra:2014aua}.

The structure of the paper is as follows. In Sec.~\ref{sec:simulations}, we provide essential details of the hydrodynamical simulations of the three $15\,M_\odot$ SN models. In Sec.~\ref{sec:evolution} the evolution of the neutrino emission characteristics of  the studied SN cases  is presented, while Sec.~\ref{sec:SASIconv} focuses on characterizing  the temporal modulations of the neutrino signal due to hydrodynamical instabilities in the presence and absence of rotation. In Sec.~\ref{sec:LESA} the effects of rotation on the LESA instability are investigated. Finally, conclusions are presented in Sec.~\ref{sec:conclusions}. Details on the method adopted in Sec.~\ref{sec:SASIconv} and employed to decouple the effects  in the neutrino signal due to SASI from the ones due to LESA are  provided in Appendix~\ref{sec:appendix}.
Animations of the evolution of the ELN  are provided as Supplemental Material.

\section{Hydrodynamical supernova simulations}\label{sec:simulations}

Details on the three adopted hydrodynamical simulations  can be found in Ref.~\cite{Summa:2017wxq}. In this Section, key distinguishing features of the SN dynamics in the three simulations are briefly outlined. 

All three simulations employ the \textsc{Prometheus-Vertex} code which includes three neutrino flavors, energy dependent and ray-by-ray-plus neutrino transport, as well as state-of-the-art modeling of the microphysics. The simulations were carried out on an axis-free Yin-Yang grid (2 degrees angular resolution for the rotating models and 4 degrees for the non-rotating one). All simulations adopt the Lattimer and Swesty nuclear equation of state with a nuclear incompressibility of 220 MeV.

\begin{figure*}
\centering
\includegraphics[width=2.\columnwidth]{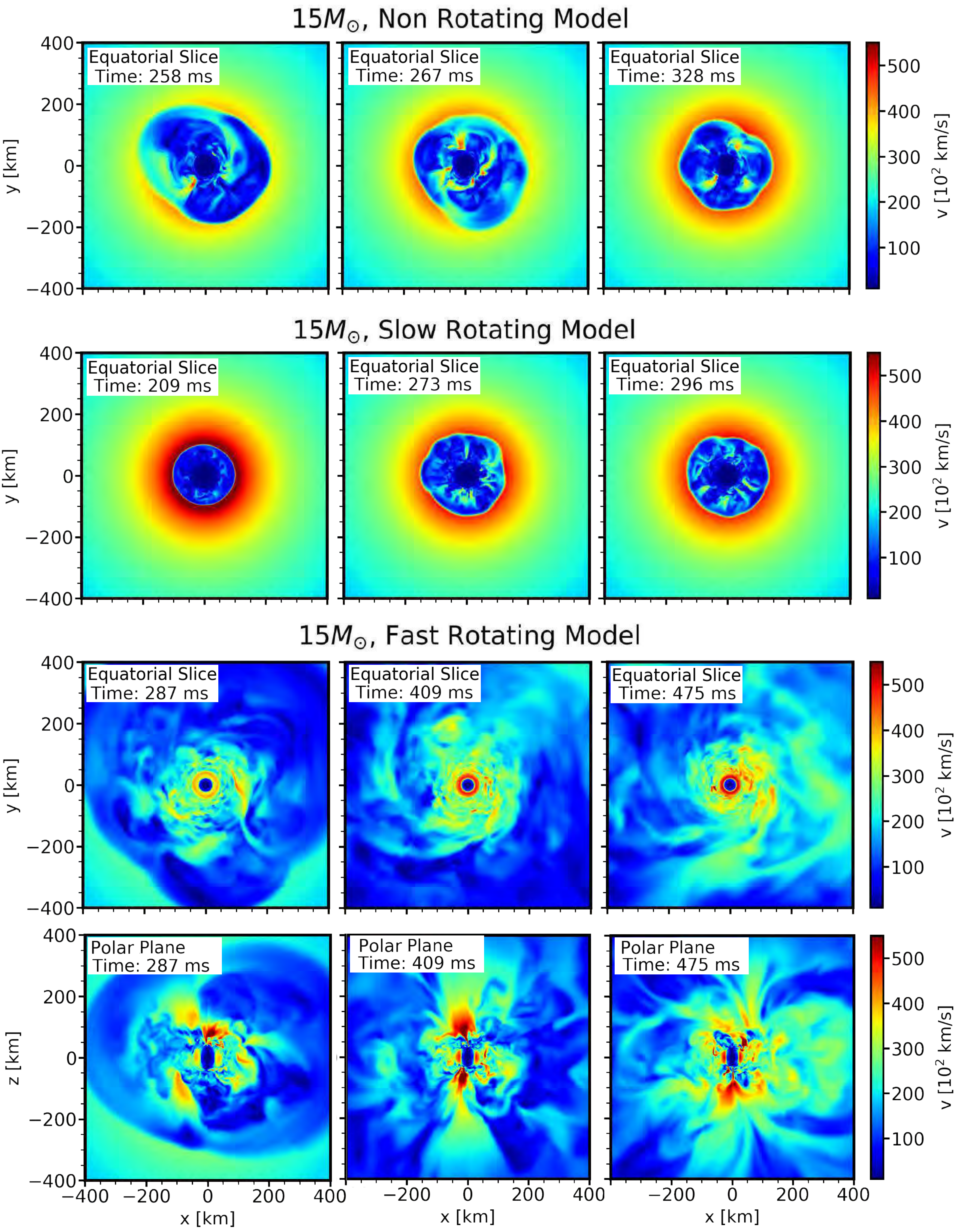}
\caption{Cross-sectional slices showing the absolute fluid velocity for the $15\,M_\odot$ non-rotating, slowly rotating, and fast rotating models (from top to bottom), at three different chosen simulation times to be motivated in Sec.~\ref{sec:LESA} (from left to right). The $z$-axis corresponds to the rotation axis, while the $x$,$y$-axes span the equatorial plane perpendicular to the rotation axis. The top panels represent the velocity distributions in the $(x,y)$ plane. The polar slices in the bottom panels represent the velocity distributions in the $(x,z)$ plane, and are oriented such that the north (south) poles are at the upper (lower) halfs of the plots. Imprints of the SASI sloshing motions are clearly visible for the non-rotating SN model, while SASI is weak in the slowly rotating model. The rapidly rotating model explodes by the aid of a strong spiral SASI mode that develops in the $(x,y)$ plane. Unsteady polar downflows of matter accreting onto the PNS affect the post-explosion dynamics of the fast-rotating model, and manifestly appear  in the polar slices of  this case.}
\label{fig:V_Evo_400}
\end{figure*}

The non-rotating model shows  SASI spiral activity in [120,250]~ms after core bounce; SASI then decays as the shock expands after the infall of the Si-SiO interface.  This can be seen in the top panels of Fig.~\ref{fig:V_Evo_400}, which show the evolution of the absolute fluid velocity in the $(x,y)$ plane corresponding to the equatorial plane of the simulation grid (i.e., the plane perpendicular to the rotation axis $z$ for the rotating models) for three selected snapshots chosen at times to be further motivated in Sec.~\ref{sec:LESA}. A global deformation of the post-shock layer in  the equatorial plane, caused by  the sloshing  motions of the SASI instability is clearly visible. 

The slowly rotating model relies on an initial angular velocity profile from stellar evolution calculations including angular momentum transport caused by  magnetic fields. The specific angular momentum  is   $6\times10^{13}$~cm$^2$~s$^{-1}$ at the Si/SiO interface prior to the core collapse;  it has a spin period of 6000 s. As evident from the panels of the second row of Fig.~\ref{fig:V_Evo_400}, this model  does not exhibit strong SASI signatures. Instead, the convective activity fostered by neutrino heating causes the formation of curl-like currents in the post-shock layer. 

The fast rotating model is the only simulation of this set which explodes successfully. It has a specific angular momentum of $2 \times 10^{16}$~cm$^2$~s$^{-1}$ at the Si/SiO interface in the pre-collapse state, corresponding to  a spin period of about 20 s. The rapid rotation triggers an early SASI spiral mode in the equatorial plane perpendicular to the rotation axis. As a result, a successful explosion occurs around 200 ms post bounce, when the in-falling Si-SiO interface reaches the shock-wave. The shock-wave is then pushed outwards and, as a consequence, neutrino heating within the gain layer is enhanced. An oblate deformation along the equatorial plane occurs which grows when the successful explosion sets in. The earliest time snapshot of the fast rotating model along the equator in Fig.~\ref{fig:V_Evo_400} displays imprints of the SASI spiral motions. The oblate deformation generated by the rapid rotation can be seen in the left bottom panel of Fig.~\ref{fig:V_Evo_400} (where the velocity distribution  is instead plotted in the $(x,z)$ plane), in the form of increased rotational velocity close to the proto-neutron star (PNS). The panels in the bottom of Fig. \ref{fig:V_Evo_400} also illustrate large-scale funnel-like structures  along the rotation axis, which correspond to unsteady polar downflows of matter accreting onto the PNS at post-explosion times.

\section{Evolution of the neutrino emission properties}\label{sec:evolution}

Neutrinos carry characteristic imprints of the hydrodynamical instabilities and the rotation dynamics which govern each SN model. In this Section, we explore these effects by comparing and contrasting the time evolution of the neutrino properties between the three SN models. 																		

\subsection{Directional neutrino emission properties}\label{sec:directional_prop}
The neutrino emission characteristics displayed throughout the rest of this work have been extracted at a radius of 500 km~\footnote{The neutrino data in the SN comoving frame have been employed, since any transformation from the comoving frame to lab frame would lead to a negligible correction.} and remapped from the Yin-Yang grid onto a standard spherical grid. The neutrino properties have then been projected to appear as they would for a distant observer at selected angular coordinates, following the procedure outlined in Appendix A of Ref.~\cite{Tamborra:2014hga} (see also Ref.~\cite{Muller:2011yi}). 
The \href{https://wwwmpa.mpa-garching.mpg.de/ccsnarchive/data/Walk2018/}{neutrino emission properties} for every angular direction as seen by a distant observer can be provided upon request for all three SN models.

\begin{figure*}
\centering
\includegraphics[width=2.\columnwidth]{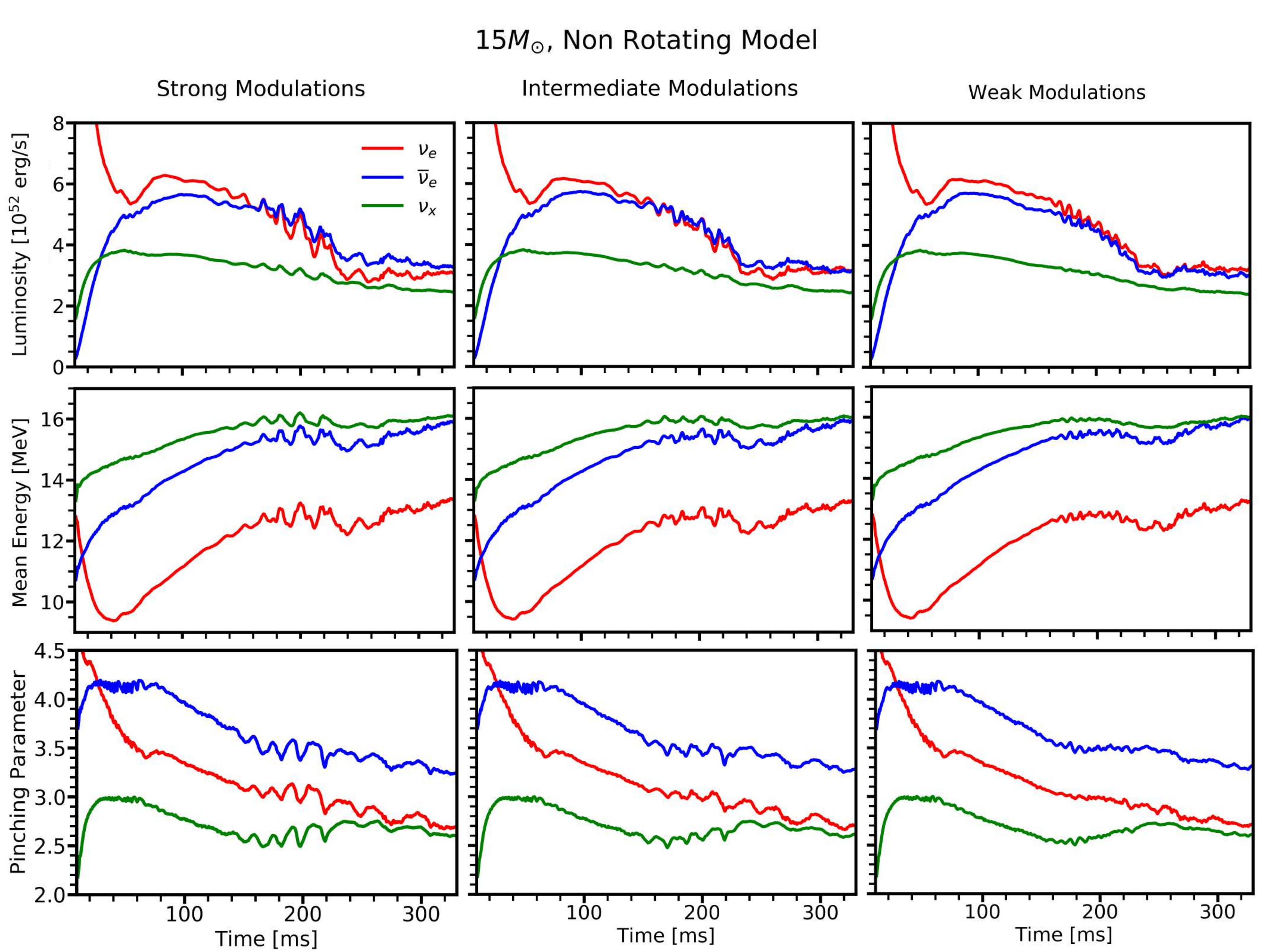}
\caption{Temporal evolution of the neutrino emission properties for the non-rotating $15\,M_\odot$ model as seen by a distant observer. For 
$\nu_e$, $\bar{\nu}_e$, and $\nu_x$ (in red, blue and green, respectively), the luminosity, mean energy, and pinching parameter are shown from top to bottom. The directions with strong, intermediate, and weak modulations have been chosen according to the directions with the maximum, intermediate, and minimum signal modulation as explained in Sec.~IV (Fig.~3) of Ref.~\cite{Walk:2018gaw}.}
\label{fig:propertiesNR}
\end{figure*}

We begin by investigating the dependence of the neutrino properties on the observer direction.  Figure~\ref{fig:propertiesNR} shows the neutrino emission characteristics of the  non-rotating $15\,M_\odot$ model along directions of strong, intermediate, and weak signal modulations. These directions have been selected by determining the total modulation   in each angular zone over the mentioned SASI time interval relative to the 4$\pi$-average.  Figure~3 of Ref.~\cite{Walk:2018gaw} shows the angular coordinates of the observer directions studied (cf.~also Sec.~IV therein); note that the selected angular directions refer to the directions where the neutrino signal exhibits the largest overall variation, see Sec.~\ref{sec:SASIconv_1} for a related discussion. For each panel, the red, blue, and green curves refer to $\nu_e$, $\bar\nu_e$, and $\nu_x$ signals respectively. We denote the non-electron flavor $\nu_{\mu,\tau}$ by $\nu_x$. For each flavor $\nu_\beta$, the panels from top to bottom show the neutrino luminosity, mean energy, and pinching factor ($\alpha_{\nu_\beta}$), respectively. The latter is defined such that~\cite{Keil:2002in,Tamborra:2017ubu} 
\begin{equation}
\label{eq:alpha}
\frac{\langle E_{\nu_\beta}^2\rangle}{\langle E_{\nu_\beta}\rangle^2} = \frac{2+\alpha_{\nu_\beta}}{1+\alpha_{\nu_\beta}}\ ,
\end{equation}
with $\langle E_{\nu_\beta} \rangle$ and $\langle E_{\nu_\beta}^2\rangle$ being the first and second energy moments.

The non-rotating $15\,M_\odot$ SN model exhibits global neutrino features similar to those observed in other existing simulations of non-rotating SN progenitors, see e.g.~Refs.~\cite{Tamborra:2013laa,Tamborra:2014hga,Tamborra:2014aua,Janka:2016fox}.
In fact, the  large-amplitude sinusoidal modulations in the neutrino signal due to SASI resemble what is displayed in Fig.~6 of Ref.~\cite{Tamborra:2014hga} for a non-rotating $27\,M_\odot$ model. However, although the properties of the $15\,M_\odot$  non-rotating model show comparable SASI signatures, this model exhibits weaker SASI than the 27 and  $20\,M_\odot$  models. Evidence for this will be presented and discussed in the following Section. 

\begin{figure*}
\centering
\includegraphics[width=2.\columnwidth]{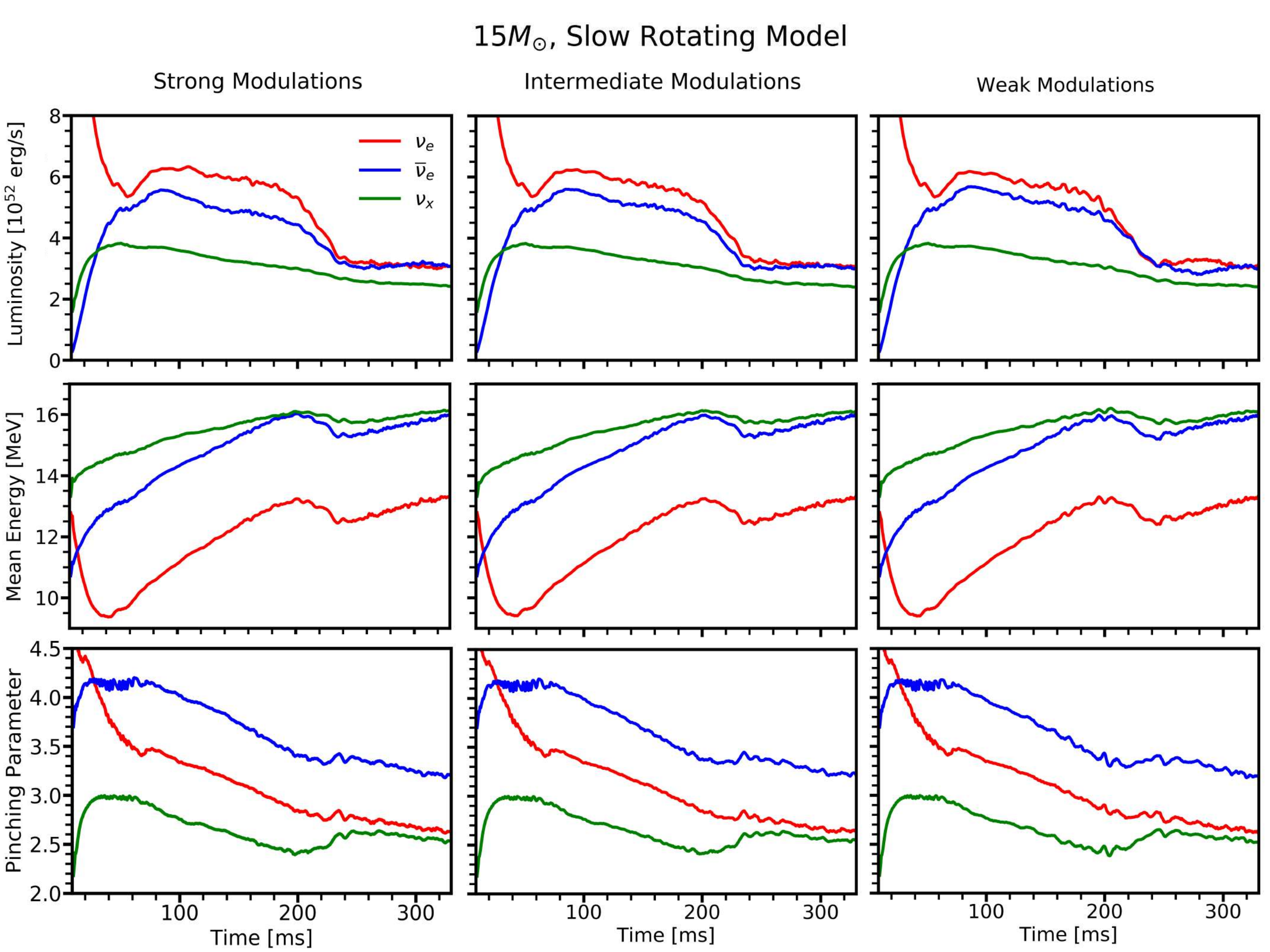}
\caption{Analogous to Fig.~\ref{fig:propertiesNR}, but for the slowly rotating $15\,M_\odot$ SN model.}
\label{fig:propertiesSR}
\end{figure*}

Figure~\ref{fig:propertiesSR} presents the neutrino emission properties for the slowly rotating model. By comparison with Fig.~\ref{fig:propertiesNR}, it is immediately apparent that the neutrino properties exhibit weaker amplitude modulations than those of the non-rotating model because of weaker SASI~\cite{Summa:2017wxq}. 
Additionally, in contrast to the non-rotating case, the amplitude of the modulations increases as the observer plane shifts from the ``strong" direction to the  direction with ``weak" modulations, an effect which is most clearly visible in the luminosity of $\nu_e$ in the SASI interval, [120,250]~ms. Remarkably, this effect, which may seem counter-intuitive at first,  is linked to the fact that 
the quadratic mean square variation, employed to select these observer directions, provides information on the overall strongest modulations, but does not necessarily select the  strongest SASI modulations, as will be discussed in Sec.~\ref{sec:SASIconv}. 
Finally,  the properties of the slowly rotating model are significantly less directionally dependent than those of the non-rotating model, again aiding the idea that global emission asymmetries are damped by rotation. 

\begin{figure*}
\centering
\includegraphics[width=2.\columnwidth]{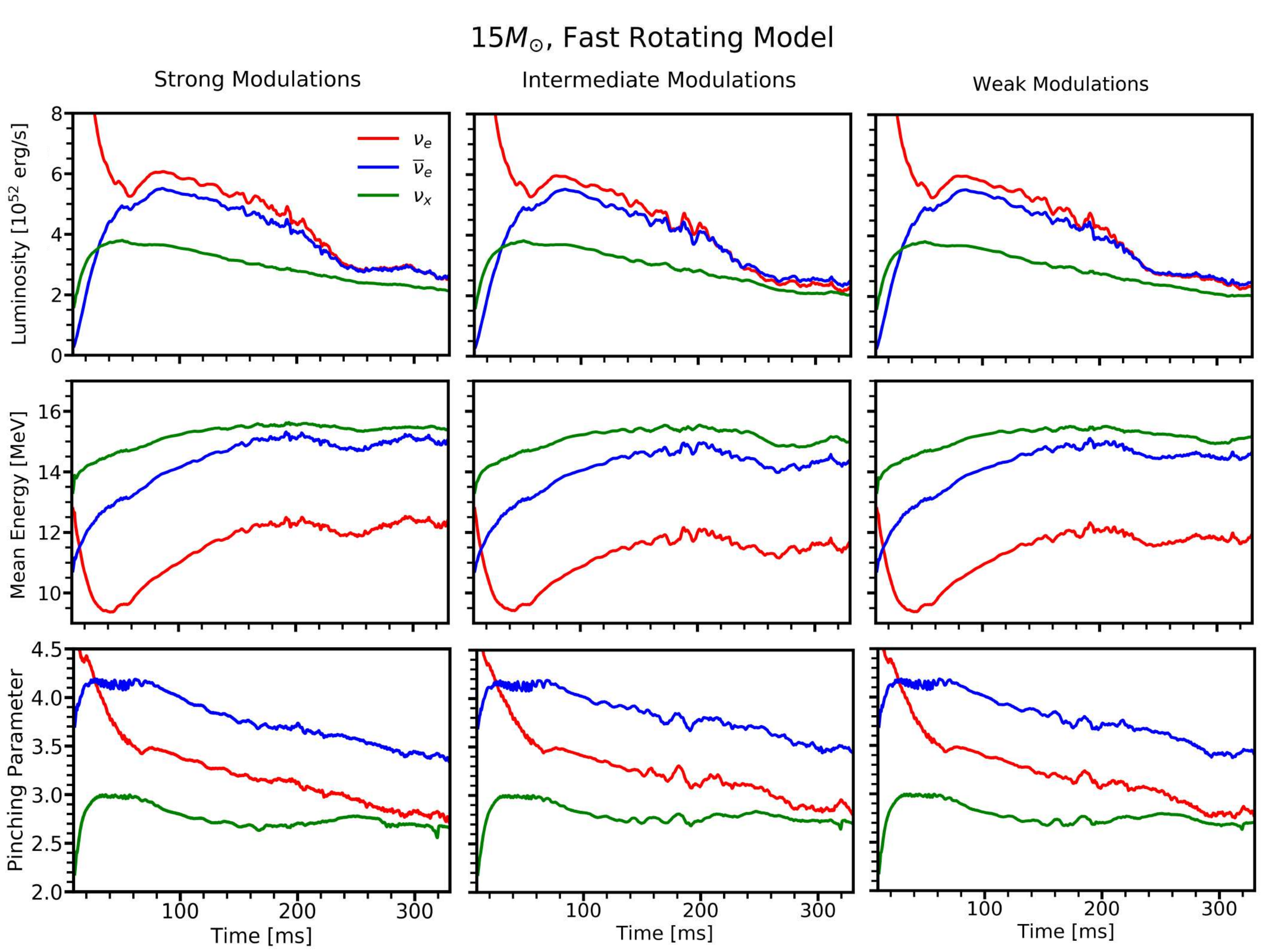}
\caption{Analogous to Fig.~\ref{fig:propertiesNR}, but for the fast rotating $15\,M_\odot$ model.}
\label{fig:propertiesFR}
\end{figure*}

The corresponding neutrino emission properties of the $15\,M_\odot$ fast rotating model are presented in Fig.~\ref{fig:propertiesFR}. Rapid rotation reduces the mass accretion rate of the newly formed neutron star, resulting in lower neutrino luminosities and mean energies compared to those observed in Figs.~\ref{fig:propertiesNR} and \ref{fig:propertiesSR}. After the explosion has set in, long-lasting, unsteady, massive polar downflows occur~\cite{Walk:2018gaw}, making the dynamics of this model considerably different from the other two cases. The polar downflows of matter induce small-scale fluctuations in the neutrino emission, which are reflected in the noise-like amplitude modulations embedded in each neutrino signal for times beyond $\sim 200$~ms. These modulations are directionally independent because matter flows between the two poles of the PNS while neutrinos are radiated (see also Fig.~1 of Ref.~\cite{Walk:2018gaw}). At the same time, the rapid rotation pulls matter out along the equatorial plane prior to the explosion and smears out the imprints of the strong SASI activity in the neutrino signal  at earlier times. These two effects both contribute to the weak directional dependence of the neutrino emission properties of the fast rotating SN model. 

\begin{figure*}
\centering
\includegraphics[width=2.\columnwidth]{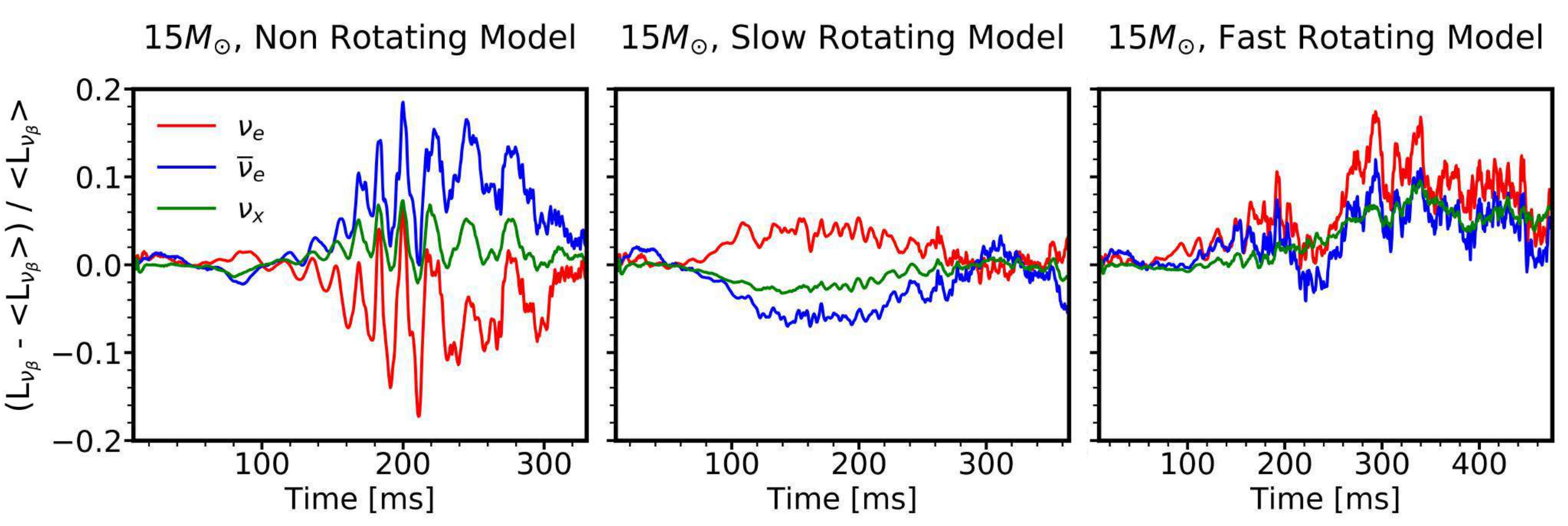}
\caption{Luminosity evolution for the non-rotating, slowly, and fast rotating models, respectively, from left to right, for the $\nu_e$, $\bar{\nu}_e$ and $\nu_x$ species as seen by a distant observer, relative to the time-dependent average luminosity over all directions ($\langle L_{\nu_\beta} \rangle$). Each luminosity is plotted along the direction of strongest modulation as defined in Sec.~IV (Fig.~3) of Ref.~\cite{Walk:2018gaw}. }
\label{fig:relatives}
\end{figure*}

Next, we investigate the effect of rotation on the neutrino luminosity of different neutrino flavors relative  to the time-dependent average luminosity over  all directions ($\langle L_{\nu_\beta} \rangle$):  
\begin{equation}
\mathcal{L}_{\nu_\beta} = \frac{L_{\nu_\beta} - \langle L_{\nu_\beta} \rangle}{\langle L_{\nu_\beta} \rangle}\ .
\end{equation}
Figure~\ref{fig:relatives} shows $\mathcal{L}_{\nu_\beta}$ as a function of time for the non-rotating, slowly, and fast rotating models, respectively from left to right. The relative luminosity is studied along the ``strong" modulation observer direction (as selected in Sec.~IV and Fig.~3 of Ref.~\cite{Walk:2018gaw}). 
 
 In the non-rotating and slowly rotating models, the relative luminosities of $\nu_e$ and $\bar{\nu}_e$ exhibit low-frequency variations in opposite directions, superimposed by correlated high-frequency SASI modulations. This implies that the SN may exhibit regions of stronger/weaker $\nu_e$ and $\bar{\nu}_e$ emission, corresponding to the LESA instability~\cite{Tamborra:2014aua}. Interestingly, the relative difference between $\nu_e$ and $\bar\nu_e$ decreases between the panels from left to right, suggesting that LESA is weakened by rotation (see Ref.~\cite{Janka:2016fox}). 
Notably, the opposite long-time trends between $\nu_e$ and $\bar\nu_e$  are not apparent in the fast rotating model, because of the effects of rapid rotation. As can be seen in Fig.~3 of Ref.~\cite{Walk:2018gaw}, the direction of strongest modulation for the fast rotating model corresponds to the north pole of the simulation grid. Thus, along this direction, all three flavors have  positive relative abundance, as matter is accreted onto the PNS near the pole. This motivates a detailed study of the LESA instability in rotating models, which will be presented in Sec.~\ref{sec:LESA}.

Lastly, as the rotational speed increases, high-frequency modulations associated with accretion variations become successively more visible in the  relative neutrino luminosity $\mathcal{L}_{\nu_\beta}$, as previously observed in Fig.~4 of Ref.~\cite{Walk:2018gaw}.  Furthermore, both $\nu_e$ and $\bar{\nu}_e$ exhibit modulations in time with greater amplitude than $\nu_x$ does, similarly to what was reported in Ref.~\cite{Tamborra:2014hga}. 

\subsection{Neutrino energy distribution}

\begin{figure}
\centering
\includegraphics[width=\columnwidth]{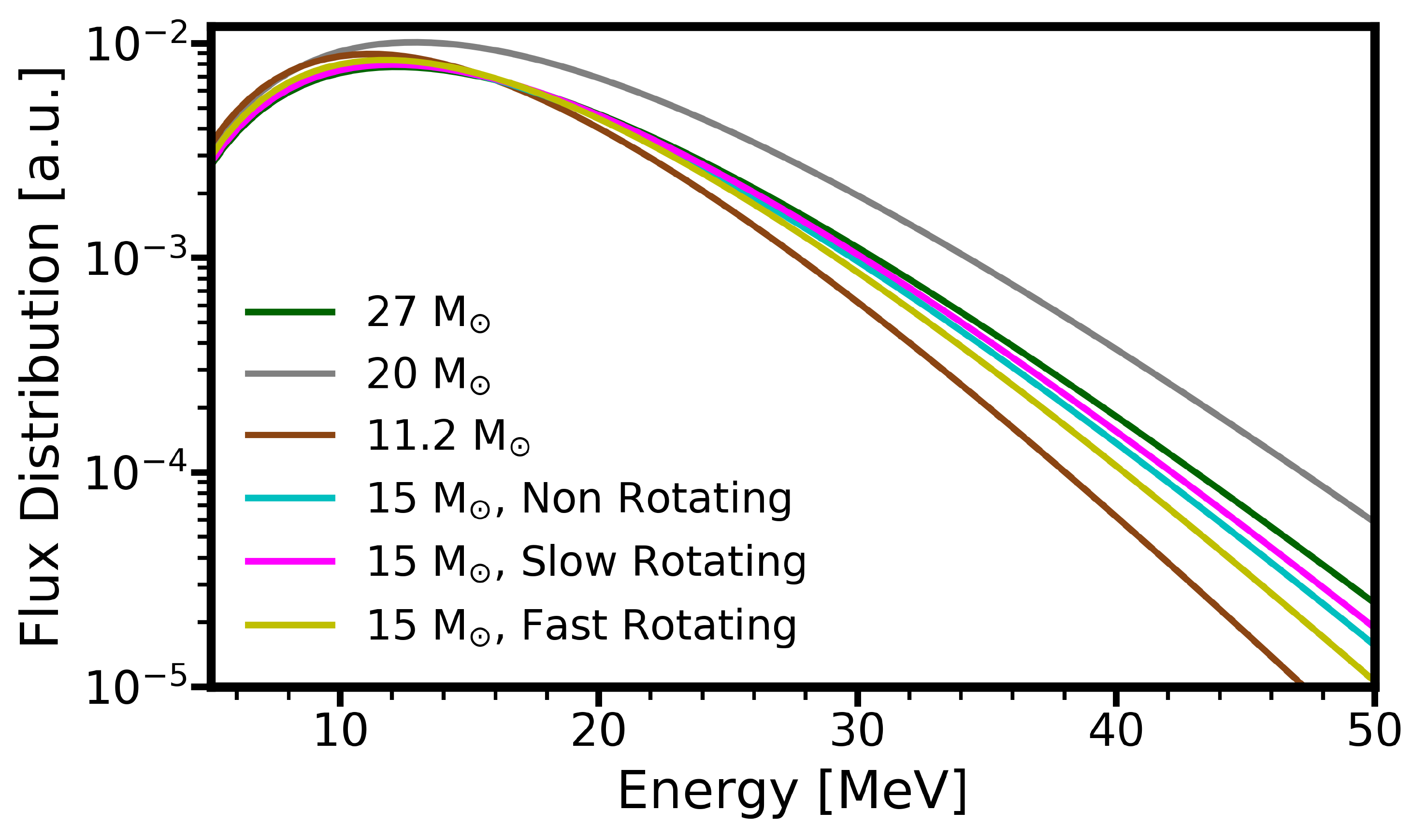}
\caption{Energy distribution for the $\bar{\nu}_e$ flavor in arbitrary units as a function of the neutrino energy  in the direction of largest signal modulation and integrated over the time window [120,250]~ms (time interval roughly corresponding to the SASI time window). The neutrino energy distribution for the fast rotating model appears to be more pinched than for the non-rotating and slowly rotating models; however, this feature due to rotation is degenerate with respect to the progenitor mass. }
\label{fig:ESpec}
\end{figure}

We now consider the effect of the SN progenitor rotation on the neutrino energy distribution. It is shown in Refs.~\cite{Keil:2002in,Tamborra:2017ubu} that the neutrino energy distribution for each flavor can be well approximated by the so-called $\alpha$-fit, given by
\begin{equation}
f_{\nu_\beta}(E) \propto \left(\frac{E}{\langle E_{\nu_\beta} \rangle}\right)^{\alpha_{\nu_\beta}} \exp^{-(\alpha_{\nu_\beta}+1) E/\langle E_{\nu_\beta} \rangle}\ .
\end{equation}
Figure~\ref{fig:ESpec} shows the normalized $\bar{\nu}_e$ energy distribution for the three $15\,M_\odot$ SN models, as well as three non-rotating models with 27, 20, and 11.2 $M_\odot$, presented in Refs.~\cite{Tamborra:2013laa,Tamborra:2014hga}, in the direction of the largest signal modulation. All neutrino spectra are integrated over the time interval [120,250]~ms. Besides decreasing the accretion luminosity emitted by neutrinos, rotation also leads to a slightly more pinched energy spectrum than in the non-rotating case; a similar trend is found for the other neutrino flavors as well. Note, however, that this cannot be used as a discriminating feature of rotation, as the dependence of the spectral pinching on the rotational speed is degenerate with variations associated with the progenitor mass.

\section{Effects of SASI and LESA emission asymmetries}\label{sec:SASIconv}

In this Section, we aim to identify the strength of SASI and LESA on the neutrino emission properties in the three $15\,M_\odot$ simulations. In particular, we are interested in exploring whether besides amplitude modulations of the neutrino signal due to SASI, the emission properties also exhibit a spread in their values as a function of the emission angle as a result of LESA. For comparison, we include  the 27, 20, and  $11.2\ M_\odot$ non-rotating SN models presented in Refs.~\cite{Tamborra:2013laa,Tamborra:2014hga}.

\subsection{Characterization of  direction-dependent variations of the neutrino properties}\label{sec:SASIconv_1}

In Refs.~\cite{Tamborra:2013laa,Tamborra:2014hga,Walk:2018gaw},  the root mean squared deviation of the neutrino luminosity in each angular direction from the time-dependent $4\pi$-average luminosity ($\sigma$, see Eq.~1 of Ref.~\cite{Tamborra:2013laa}) is adopted
to estimate the variation in the neutrino emission at each point on the emission sphere. For the $27$ and $20\,M_\odot$ models, where SASI is the dominant hydrodynamical instability compared to the other  instabilities, the $\sigma$ parameter was found to be a good quantity to pinpoint the signal modulations coming from SASI directly.

In addition to the amplitude modulations in the neutrino signal arising from SASI, other  phenomena such as LESA (discussed in detail in Sec.~\ref{sec:LESA}) may be at play and may lead to variations of the neutrino properties between different directions. In these cases,  the $\sigma$ parameter may not be an optimal choice to determine directions of strong modulations coming from these different instabilities (see also discussion in Sec.~\ref{sec:directional_prop}). Thus,  we introduce two new parameters which help to determine  the neutrino luminosity variations due to SASI or LESA  exclusively. For simplicity, we focus on the $\bar{\nu}_e$ luminosity, but a similar trend is found for the other neutrino flavors as well. 

For each angular direction $(\theta,\phi)$, we begin by defining a time-dependent luminosity $\tilde{L}_{{\bar{\nu}_e}}(\theta, \phi)$ as the ``mid-line" luminosity found by determining  the local maxima and minima of the luminosity signal, and interpolating halfway between them, see Appendix~\ref{sec:appendix} for more details. We then define 
\begin{eqnarray}
\Delta &=& \frac{1}{T} \int_{t_1}^{t_2} dt \left|\frac{L_{\bar{\nu}_e}(\theta,\phi)-\tilde{L}_{{\bar{\nu}_e}}(\theta,\phi)}{\tilde{L}_{\bar{\nu}_e}(\theta,\phi)}\right|\ \   \mathrm{,}\nonumber \\ \rho &=& \frac{1}{T} \int_{t_1}^{t_2} dt \left|\frac{\tilde{L}_{\bar{\nu}_e}(\theta,\phi)-\langle\tilde{L}_{\bar{\nu}_e}\rangle}{\langle\tilde{L}_{\bar{\nu}_e}\rangle}\right|\ ,
\label{eq:deltarho}
\end{eqnarray}
where $T = (t_2 - t_1)$ and $\langle\tilde{L}_{\bar{\nu}_e}\rangle$ again denotes the time-dependent average of the $\bar{\nu}_e$ luminosity over all directions. For each angular direction, the $\Delta$ parameter gives the mean absolute  deviation of the luminosity from the corresponding mid-line luminosity in that same angular direction. This estimates the average variation of the signal amplitude, thus quantifying the effects due to the SASI instability. Contrarily, the $\rho$ parameter gives the mean absolute deviation of the mid-line luminosity at each angular zone from the $4\pi$-average of the mid-line luminosity. This quantifies the spread in luminosities over all angular emission directions, which is caused by LESA. The parameter $\Delta$ can be considered as a measure of the high-frequency neutrino-emission modulations that are characteristic of SASI and convective accretion variations, whereas $\rho$ measures the large-scale emission asymmetries associated with LESA.

\begin{figure*}
\centering
\includegraphics[width=1.35\columnwidth]{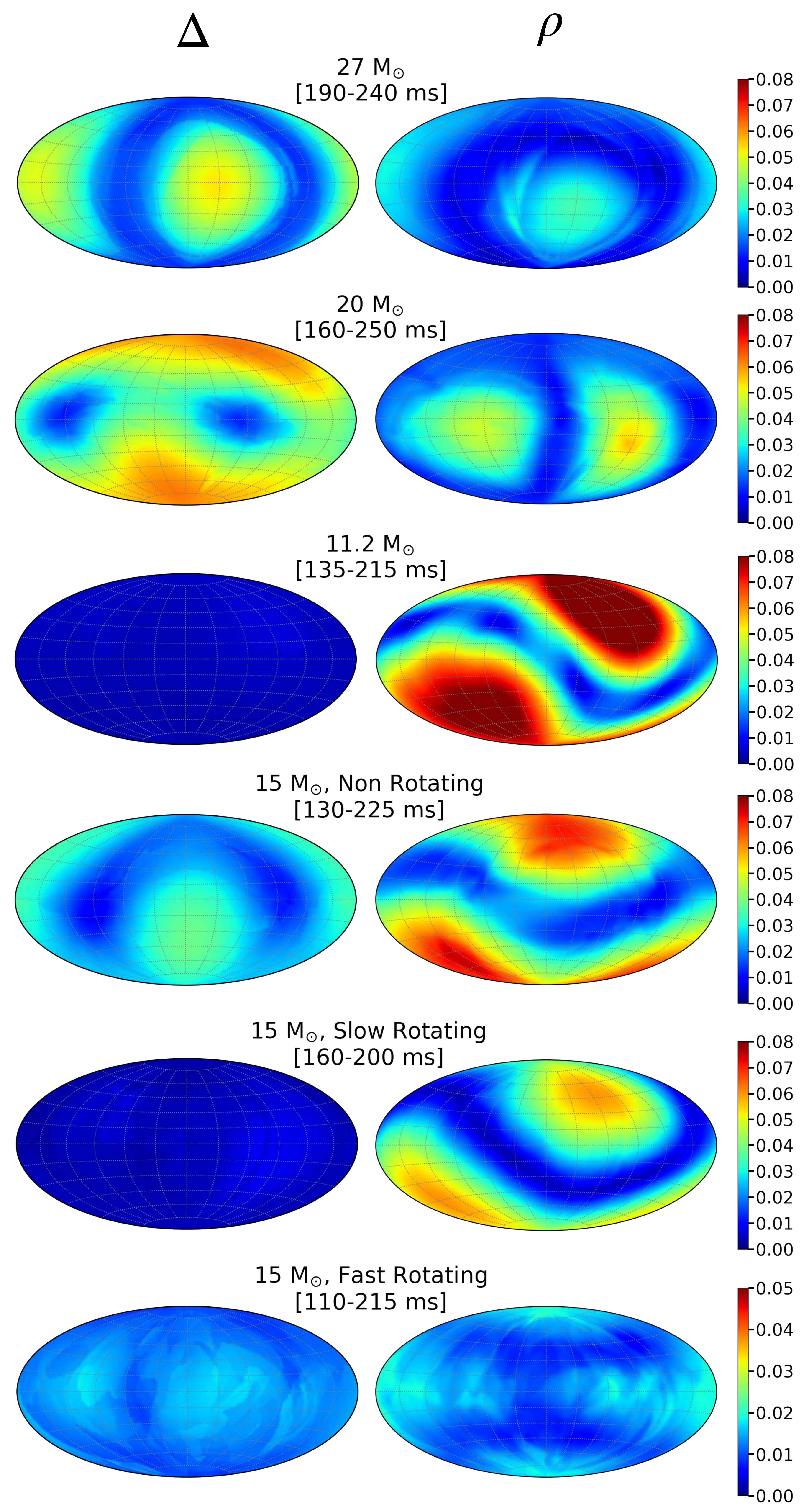}
\caption{Molleweide maps of the $\Delta$ (left) and $\rho$ (right) modulations as defined in Eq.~\ref{eq:deltarho} for the $27, 20, 11.2$, and three $15\ M_\odot$ SN models from top to bottom.  The chosen time intervals lie within the SASI episode for the models that have SASI ([120,250] ms for the 27, 15, and 11.2 $M_\odot$ models and [150,330] ms for the 20 $M_\odot$ model), and correspond to the interval in which the signal amplitude varies maximally so that the mid-line interpolation can be found to the highest degree of accuracy. The boundaries of the color bar  have been changed for the $15\ M_\odot$ fast rotating model in order to enhance the smaller $\Delta$ and $\rho$ modulations in this model compared to the others. The $27$ and $20\ M_\odot$ SN models show the largest $\Delta$ modulations, these models having the strongest SASI activity. The $11.2\ M_\odot$ model has no modulation in $\Delta$, and very strong modulations in $\rho$,  LESA being prominent in this model.}
\label{fig:Deltarho}
\end{figure*}

Figure~\ref{fig:Deltarho} shows Molleweide maps of $\Delta$ (left) and $\rho$ (right) for the $\bar{\nu}_e$ luminosity. The interval $[t_1, t_2]$ varies between  the  models as indicated in the figure; the time interval $[t_1, t_2]$ has been chosen to minimize the error in the averaging procedure to determine the mid-line luminosity that, as discussed in Appendix~\ref{sec:appendix},  lies within the time-interval where SASI occurs.  As expected, the $27$ and $20\ M_\odot$ SN models have the largest $\Delta$ modulations as these models have the strongest SASI activity. Although they exhibit some $\rho$ modulations and LESA is present, SASI is clearly the dominant instability affecting the neutrino signal. On the contrary, the $11.2\ M_\odot$ model has no modulation in $\Delta$, and very strong variations in $\rho$; in fact, it is completely dominated by convection and develops a prominent LESA over time~\cite{Tamborra:2014aua}.
 
Interestingly, the $15\ M_\odot$ non-rotating model shows the second highest $\rho$ modulations. It also shows weak SASI activity, indicating that this model experiences effects from both SASI and LESA. In fact, when considering just the four non-rotating models (27, 20, 15, and 11.2 $M_\odot$), a clear trend with respect to SASI dominance as a function of the progenitor mass can be identified. In decreasing SN progenitor mass, the models go from having almost completely SASI dominated activity to an absence of SASI activity, with the 20 and 15 $M_{\odot}$ models falling in between. 

When studying the two $15\ M_\odot$ rotating cases and comparing them to the $15\ M_\odot$ non-rotating model, we see that both the $\Delta$ and $\rho$ modulations decrease with rotational speed. Notably, the fast rotating model shows slightly larger modulations in $\Delta$ relative to the slowly rotating model because of variations induced in the neutrino signal by the post-explosion polar downflows of matter. The disappearance of $\Delta$ modulations between the non-rotating and slowly rotating models is, again, clear evidence that rotation smears out global SASI neutrino-emission asymmetries, despite the fact that SASI spiral motions are very strong in the fast rotating model. 

Rotation decreases the spread in luminosities relative to the $4\pi$-average as well (see also Appendix~\ref{sec:appendix}). This suggests that rotation also disfavors the formation of regions of asymmetric neutrino emission coming from LESA across the surface of the progenitor; this effect was also apparent from Fig.~\ref{fig:relatives}; it  fully agrees  with the results of LESA presented in Ref.~\cite{Janka:2016fox}.

\subsection{Multipole analysis of the neutrino luminosity}\label{sec:multipoles}

As was apparent from the figures in Sec.~\ref{sec:evolution}, with increasing rotational velocity,  the neutrino-emission   modulations due to SASI and/or LESA are weakened. In order to investigate this in more detail, we decompose the luminosity  $L_{\bar\nu_e}$ into spherical harmonics at the source (i.e., before observer projections) following Refs.~\cite{Janka:2016fox,Burrows:2012yk}, such that
\begin{equation}
A_l = \sqrt{\sum_{m=-l}^l a_{lm}^2}\ ,
\end{equation}
where
\begin{equation}
a_{lm}=\frac{(-1)^{|m|}}{\sqrt{4\pi (2 l +1)}} \int L_{\bar\nu_e}(\theta,\phi) Y^m_l(\theta,\phi) d\Omega \ .
\end{equation}
The orthonormal harmonic basis functions are
\begin{equation}
Y^m_l(\theta,\phi)=\begin{cases}
\sqrt{2} N_l^m P_l^m(\cos\theta) \cos(m\phi) &m>0\\
N_l^0 P_l^0(\cos\theta) &m=0\\
\sqrt{2} N_l^{|m|} P_l^{|m|}(\cos\theta) \sin(|m|\phi)\quad&m<0 
\end{cases}\,,
\end{equation}  
with
\begin{equation}
N_l^m = \sqrt{\frac{2 l +1}{4\pi} \frac{(l-m)!}{(l+m)!}}\ ,
\end{equation}
where $P_l^m(\cos\theta)$ are the Legendre polynomials. To be consistent with the normalization used in Ref.~\cite{Tamborra:2014aua}, we normalize the amplitude $A_l$ by $(2 l +1)$. According to this formalism, the coefficients $a_{11}$, $a_{1-1}$ and $a_{10}$ correspond to the average Cartesian coordinates of $L_{\bar\nu_e}$ ($x$, $y$ and $z$, respectively).

\begin{figure*}
\centering
\includegraphics[width=2.\columnwidth]{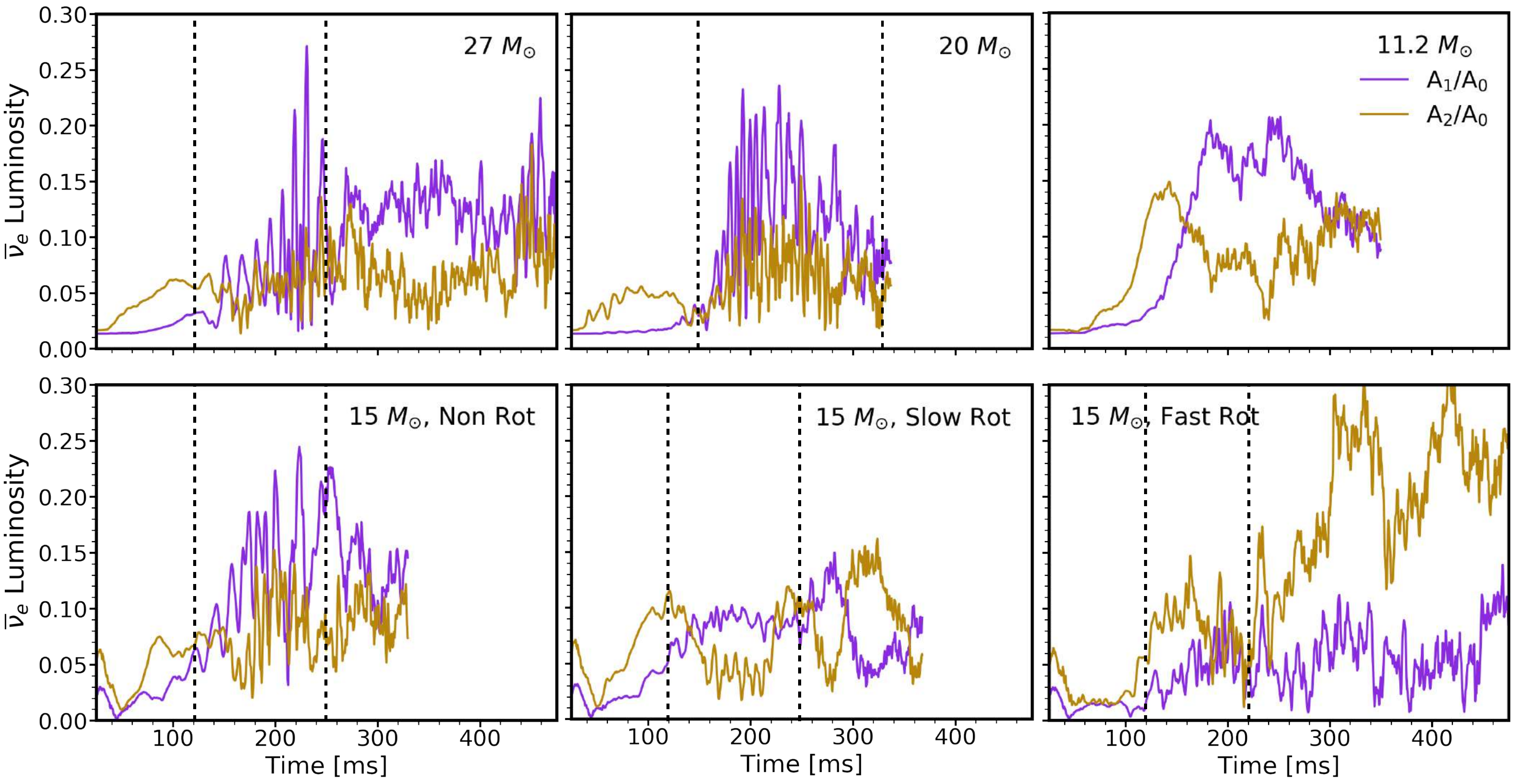}
\caption{Time evolution of the dipole and quadrupole strength of $L_{\bar\nu_e}$ normalized to the monopole for the $27, 20, 11.2$ and the three $15\ M_\odot$ SN models. The dashed vertical lines mark the SASI interval for each SN model. The $L_{\bar\nu_e}$ quadrupole dominates for all models until the beginning of the SASI phase. At later times, the dipole takes over in all of the non-rotating models. Notably, the post-explosion polar accretion flows in the fast rotating model are responsible for the dominance of the quadrupolar mode and its growth at later times. }
\label{LESA_SASI_Mulitpoles}
\end{figure*}

Figure~\ref{LESA_SASI_Mulitpoles} shows the $L_{\bar\nu_e}$ dipole and  quadrupole normalized to the monopole (i.e., $A_{l=1,2}/A_0$) as a function of time. The $L_{\bar\nu_e}$ quadrupole ($l=2$) dominates for all models until $\sim 120$~ms ($\sim 150$~ms for the 20 M$_{\odot}$ model), roughly corresponding to the beginning of the SASI phase. This is in agreement with the findings of Ref.~\cite{Glas:2018vcs} which discusses that high $l$ modes (starting with $l=4$) develop earlier than the low $l$ ones. 

At later times, the dipole takes over in all of the non-rotating models.  The slowly rotating model experiences periods when the dipole or the quadrupole is dominant;
in particular, the slowly rotating model has a dominant dipole moment exactly during the SASI episode because of the  SASI shock motions inducing dipolar asymmetries in the mass accretion of the PNS, and thus in the $\nu_e$ and $\bar{\nu}_e$ emission. 
 In the case of the rapidly rotating model, instead,  the two persistent polar accretion flows are responsible for the dominance of the $L_{\bar\nu_e}$ quadrupole mode in the neutrino emission. This is a clear feature associated with  the SN rotation.  

The $11.2~M_\odot$ model is a peculiar model because it does not have any SASI activity. In this case, the prominent dipole emission mode visible in Fig.~\ref{LESA_SASI_Mulitpoles} is a pure consequence of LESA (while in all other models the LESA and SASI emission asymmetries are superimposed, both contributing to the dipole). Notably, while the SASI dipoles of all neutrino species are correlated, the LESA dipoles of $\nu_e$ and $\bar{\nu}_e$ are anti-correlated and exhibit only long-time variations instead of the higher frequency modulations of the SASI dipole emission~\cite{Tamborra:2014aua,Tamborra:2014hga}, see Fig.~\ref{fig:relatives} for both kinds of correlated and anti-correlated behavior between $\nu_e$ and $\bar{\nu}_e$.

\section{LESA in rotating supernova models}\label{sec:LESA}

In Ref.~\cite{Tamborra:2014aua}  a large hemispheric asymmetry in the ELN emission (LESA) from the newly formed PNS  was discovered in a set of 3D SN simulations based on the ray-by-ray plus approximation. It was found that the dipole amplitude and its direction exhibit a stationary or slowly migrating behavior over time intervals of several hundreds of milliseconds (see Fig.~1 of Ref.~\cite{Tamborra:2014aua}). These findings have been confirmed in Ref.~\cite{Janka:2016fox} in which  a larger number of SN models is examined. Independent confirmation of LESA has been presented  in Refs.~\cite{OConnor:2018tuw,Vartanyan:2018iah,Glas:2018vcs} recently. These groups have applied a multi-dimensional transport scheme, providing evidence that LESA is not a numerical artefact arising from the ray-by-ray approximation. In Ref.~\cite{Glas:2018vcs}, LESA has been witnessed in a set of eight self-consistent simulations with fully multi-dimensional and ray-by-ray transport. 

As previously seen in Fig.~\ref{fig:relatives}, there is a strong anti-correlation between the ${\nu}_e$ and  $\bar{\nu}_e$ luminosities in a given angular direction for  the non-rotating and slowly rotating models. The magnitude of this anti-correlation seems to decrease with rotation. In this Section, we intend to further investigate how LESA affects the neutrino emission properties in the presence of rotation.  

\subsection{LESA characterization and multipole analysis}\label{sec:LESAmultipoles}

\begin{figure*}
\centering
\includegraphics[width=2.\columnwidth]{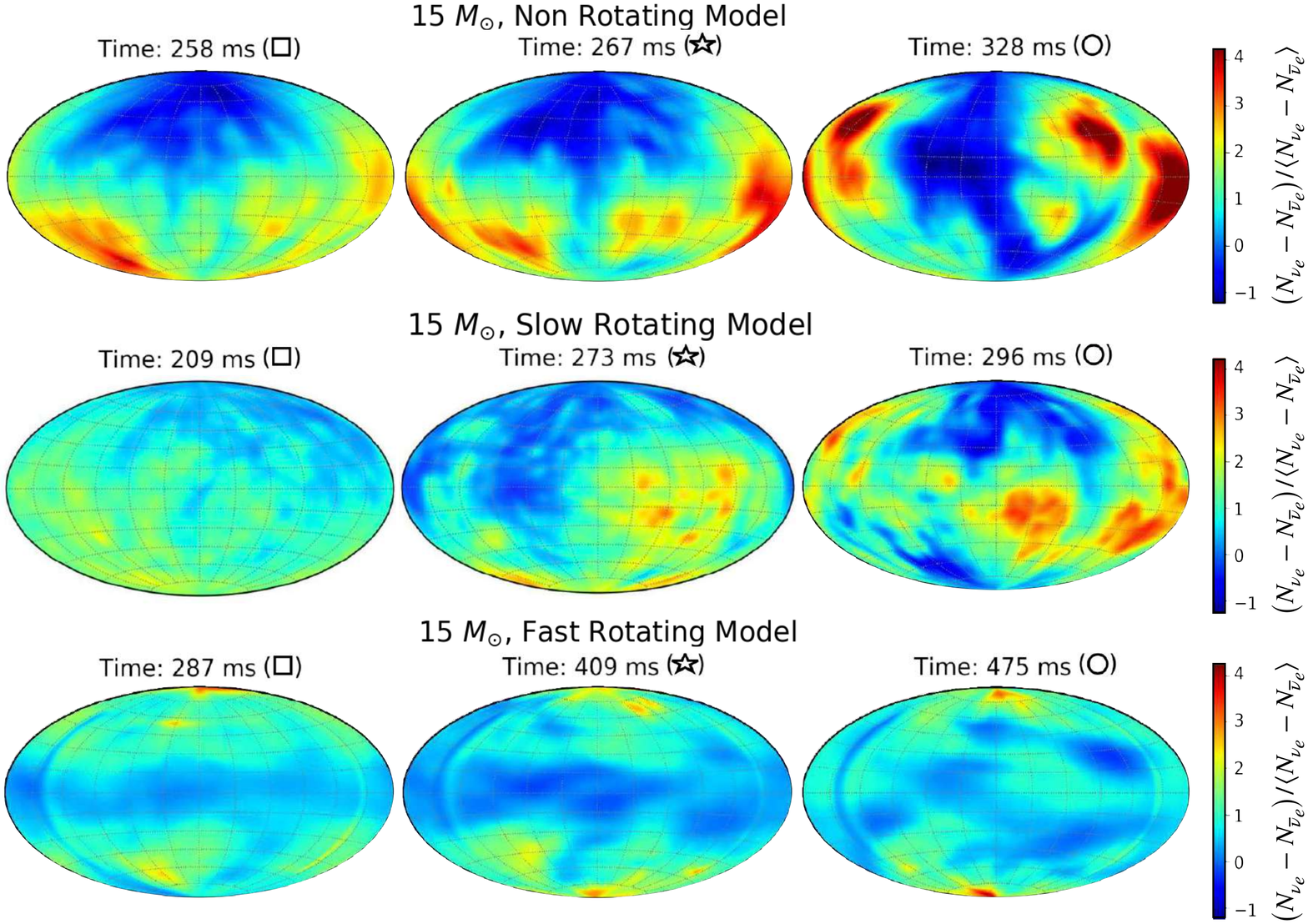}
\caption{Snapshots of the ELN flux relative to its $4\pi$-average on the emission surface, $(N_{\nu_e}-N_{\bar{\nu}_e})/\langle N_{\nu_e}-N_{\bar{\nu}_e} \rangle$, of the non-rotating, slowly rotating, and fast rotating  $15\ M_\odot$ SN models (from top to bottom) respectively. While LESA is visible for the non-rotating model, its evidence becomes weaker as the rotational speed increases. The markers $\Box$, $\star$, $\circ$ are assigned to indicate the first, second, and third time snapshots, respectively.}
\label{fig:LESA_sanpshots}
\end{figure*}

Figure~\ref{fig:LESA_sanpshots} shows a set of snapshots of the ELN flux  relative to its $4\pi$-average ($(N_{\nu_e}-N_{\bar{\nu}_e})/\langle N_{\nu_e}-N_{\bar{\nu}_e} \rangle$) for each SN model at the source (i.e., before projecting the emitted number flux to a distant observer) to illustrate the evolution of the ELN  over time. A strong ELN asymmetry develops in the non-rotating SN model, and the relative strength of this asymmetry decreases with the rotational velocity; whereas only slight asymmetries form at the poles (i.e., perpendicularly to the rotational plane) in the fast rotating model. \href{https://wwwmpa.mpa-garching.mpg.de/ccsnarchive/data/Walk2018/}{Animations of the ELN flux evolution} on Molleweide maps of the emission surface are provided as Supplemental Material  of this paper. 

\begin{figure*}
\centering
\includegraphics[width=2.\columnwidth]{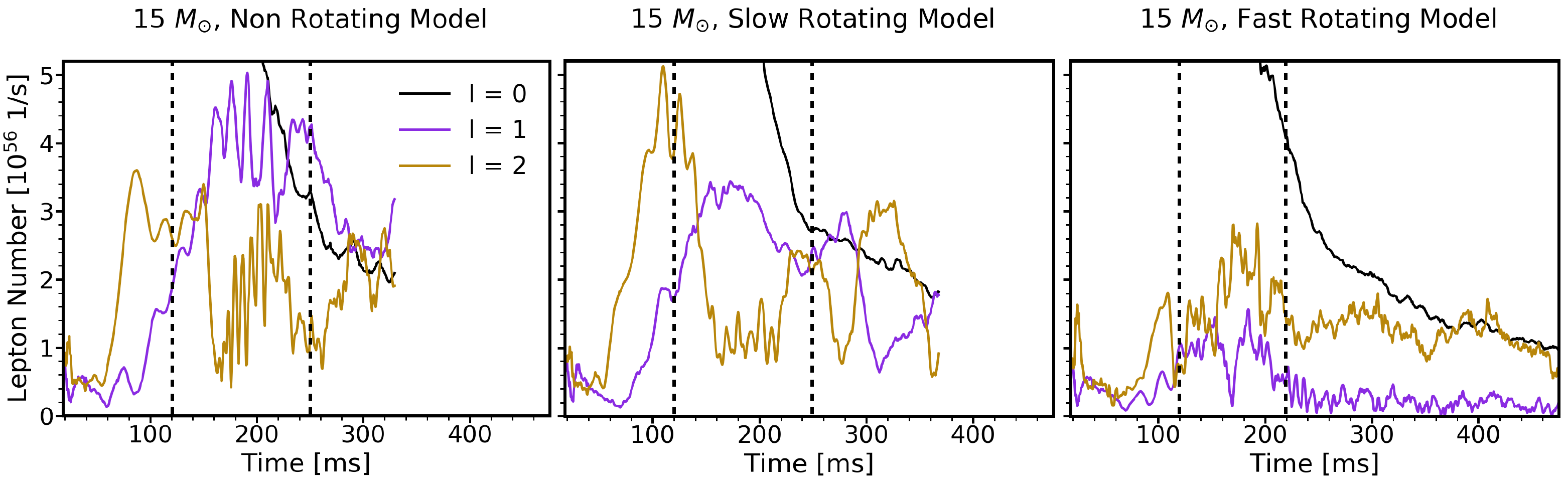}
\caption{Time evolution of the monopole (black), dipole (purple), and quadrupole (gold) amplitudes of the ELN emission for the non-rotating, slowly rotating, and fast rotating $15\,M_\odot$ models from left to right, respectively. The dashed vertical lines mark the SASI interval for each SN model. The monopole evolution is similar for all SN models, whereas the dipole amplitude decreases as the rotational velocity increases. A dominance of the quadrupole  is observable for the fast rotating model at all times.}
\label{fig:ELNdipole}
\end{figure*}

In order to quantify the time evolution of the ELN dipole, we adopt a decomposition in spherical harmonics of the ELN (see Sec.~\ref{sec:multipoles} for details). Figure~\ref{fig:ELNdipole} shows the time evolution of the ELN monopole, dipole and quadrupole ($A_{\mathrm{ELN}, 0}$, $A_{\mathrm{ELN}, 1}$ and $A_{\mathrm{ELN}, 2}$, respectively). The ELN dipole is clearly weakened by the progenitor rotation~\cite{Janka:2016fox}, while the monopole is similar in  all three models, as reported in Ref.~\cite{Tamborra:2014aua}. Moreover, as the rotational velocity increases, the relative size of the quadrupole grows; the latter fully overtakes the dipole in the fast rotating model. Thus, the ELN multipole decomposition shows a similar pattern to the one shown in Fig.~\ref{LESA_SASI_Mulitpoles}, supporting our conjecture that rotation inhibits the development of LESA. 

\begin{figure*}
\includegraphics[width=2.\columnwidth]{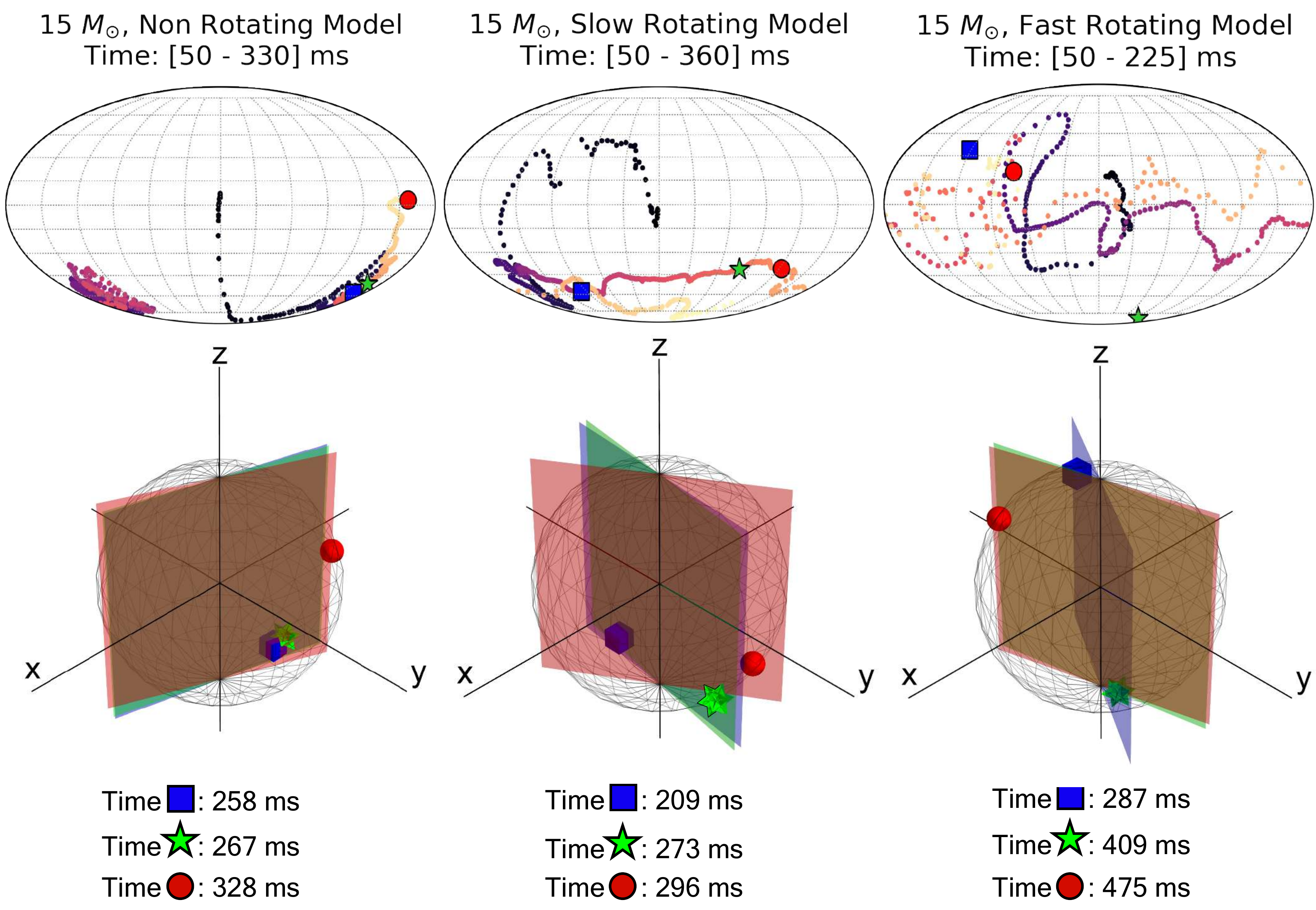}
\caption{Top panels: Evolution of the LESA (or ELN) dipole axis for the non-rotating, slowly rotating, and fast rotating $15\,M_\odot$ SN models (from left to right), respectively. The dots mark the path of the positive ELN/LESA  dipole direction. The color hues become lighter as time increases. The ELN/LESA  axis stabilizes in the bottom hemisphere for the non-rotating model, but is unstable for the fast rotating model. The markers $\Box$, $\star$, $\circ$ are assigned to indicate the positive ELN/LESA dipole moment in the  first, second, and third time snapshots as defined in Fig. \ref{fig:LESA_sanpshots} respectively. Bottom panels: ELN/LESA  planes for the non-rotating, slowly rotating, and fast rotating $15\,M_\odot$ SN models (from left to right), respectively, at the snapshot times defined in Fig. \ref{fig:LESA_sanpshots}.}
\label{fig:ELNaxis}
\end{figure*}

We also examine the time evolution of the ELN dipole axis (corresponding to the LESA dipole in the presence of the LESA instability) in each SN model. The top panels of Fig.~\ref{fig:ELNaxis} track the ELN/LESA dipole direction over a selected time interval. To guide the eye and facilitate a comparison, the location of the positive ELN/LESA dipole direction at the times of the three snapshots in Fig. \ref{fig:LESA_sanpshots} are marked on each map. 

In the non-rotating model, the ELN/LESA dipole stabilizes in the bottom hemisphere in the interval during which the power of the dipole moment is dominant over the quadrupole moment ([150,300] ms from Fig.~\ref{fig:ELNdipole}); this is a general trend for non-rotating SN models found in Ref.~\cite{Tamborra:2014aua}. For earlier and later times (indicated in Fig.~\ref{fig:ELNaxis} by the darkest and lightest colors, respectively), the quadrupole moment is dominant, and a  ``sweeping'' of the dipole direction over the SN surface is observed. 

\begin{figure*}
\includegraphics[width=2.\columnwidth]{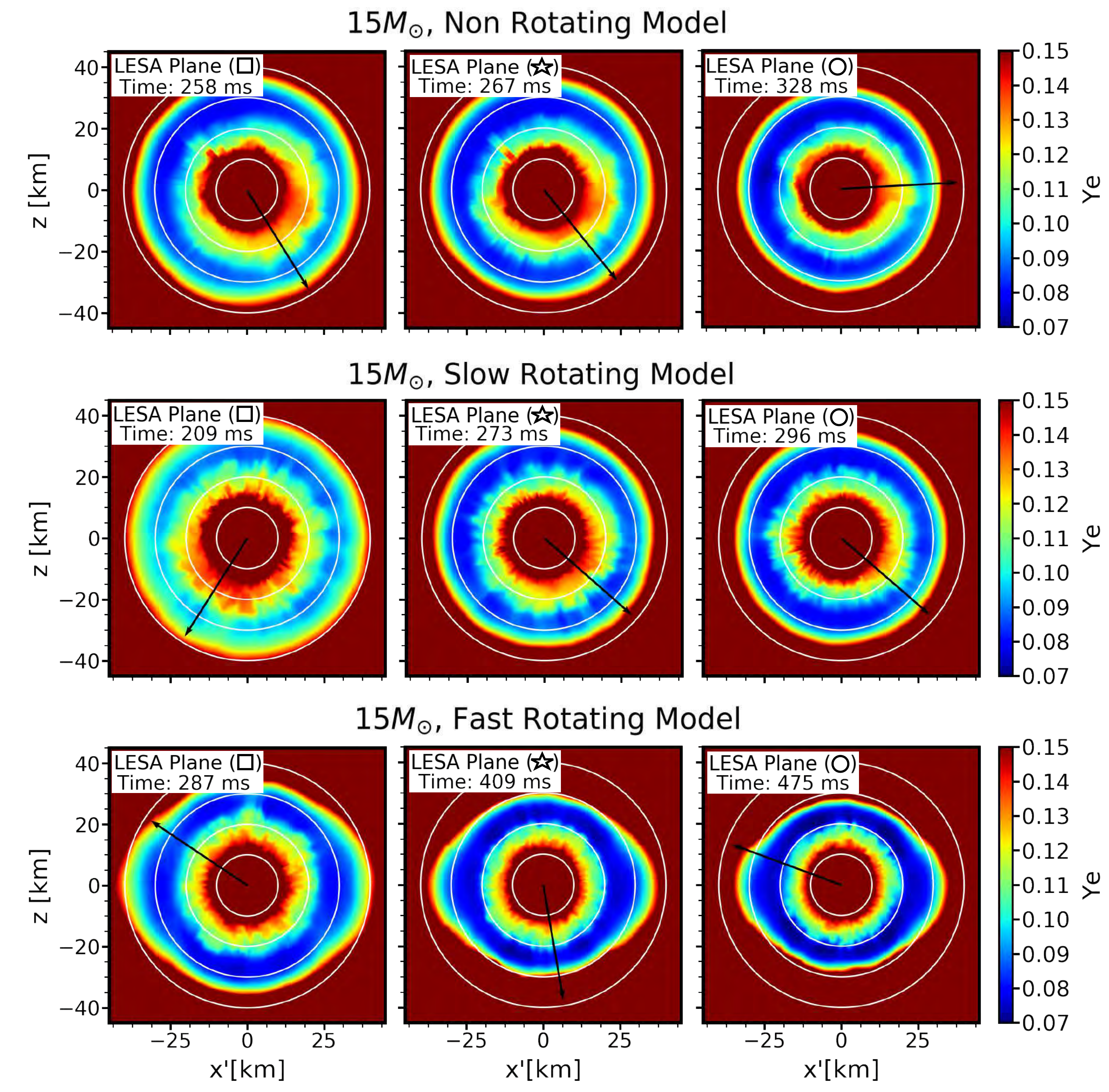}
\caption{Slices showing the spatial distribution of the electron fraction, $Y_e$, inside the PNS, taken at the times of the snapshots in Fig.~\ref{fig:LESA_sanpshots} for the non-rotating, slowly rotating, and fast rotating models from top to bottom, respectively. Each cut is taken along the plane containing the ELN dipole moment vector and the rotational axis ($z$-axis), see Fig.~\ref{fig:ELNaxis}. The plots are oriented such that the north and south poles are at the top and bottom, respectively. The black arrows point toward the positive ELN/LESA dipole moment as defined in Fig.~\ref{fig:ELNaxis} by the $\Box$, $\star$, and $\circ$ markers for each time snapshot.  The white concentric contour lines indicate radii at every 10 km to guide the eye.}
\label{fig:Yeevo}
\end{figure*}

In the slowly rotating model, we see that the ELN/LESA axis remains in the lower hemisphere, but is dragged over the radiating surface on a path around  the rotation axis. This is enhanced in the fast rotating model, where LESA appears to be very weak and the increased rotational speed favors an emission dipole near the equatorial plane.  In the fast rotating model, the ELN dipole is comparable or smaller to the quadrupole during the SASI phase prior to explosion; we suspect that the weak dipole emission is a consequence of the SASI spiral mode and its associated asymmetric PNS accretion, and not to the LESA instability. The interpretation that this phenomenon is {\emph{not}} identical with the generic appearance of LESA is supported by Fig.~\ref{fig:Yeevo}, where it is shown that the fast rotating model does not exhibit any sign of a hemispheric electron fraction ($Y_e$) asymmetry inside  the convective PNS layer. 

The bottom panels of Fig.~\ref{fig:ELNaxis} show the location of the ELN/LESA planes at the times of the three snapshots in Fig. \ref{fig:LESA_sanpshots}. The ELN/LESA plane has been defined as the plane spun by the rotation axis ($z$ axis) and the ELN/LESA dipole axis. The red, green, and blue planes contain the ELN/LESA dipole moment vector, and the rotation axis for the first, second, and third time snapshot, respectively. The location of the positive dipole direction at each time (as indicated on the Molleweide maps in the top panel of the figure) is marked on each plane. For the sake of simplicity, we denote this plane by $(x^\prime,z)$, where the $x^\prime$ axis is obtained by projecting the ELN/LESA dipole axis on the $(x,y)$ plane of the grid used for the analysis.

It was shown in Refs.~\cite{Tamborra:2014aua,Janka:2016fox} that LESA is associated with the formation of a deleptonized region in the direction of the minimal ELN flux, corresponding to a lower electron fraction ($Y_e$).
Figure~\ref{fig:Yeevo} shows cuts of the spatial distribution of $Y_e$. In the  $(x^\prime,z)$ planes  containing the ELN/LESA dipole axis (see Fig. \ref{fig:ELNaxis}), and the rotation axis. The plots are oriented such that the north and south poles are at the top and bottom of each plot respectively.  One can see that the non-rotating model develops a clear asymmetry in $Y_e$, similar to what was observed in Refs.~\cite{Tamborra:2014aua,Janka:2016fox}. On the other hand, we find that the $Y_e$ distribution becomes more symmetrical as the SN rotational velocity increases, implying that  the development of LESA is inhibited.

\subsection{Development of LESA in rotating models}\label{sec:LESAevolution}

As pointed out in Refs.~\cite{2001LNP...578..333J},  the strength of convection in rotating and non-rotating SN models is different. In 2D simulations of rotating SNe, it was found that  convective activity is confined to the region  close to the equator, whereas it involves a whole spherical shell in the corresponding non-rotating model (see Figs.~14 and 11 of Ref.~\cite{2001LNP...578..333J}). Since the development of LESA turned out to be strongly connected to the convective activity within the PNS~\cite{Tamborra:2014aua,Janka:2016fox,Glas:2018vcs,Powell:2018isq}, we now investigate  whether the suppression of LESA  in the rotating models originates from a weakening of the convective overturn within the PNS due to rotation.

\begin{figure*}
\includegraphics[width=2.\columnwidth]{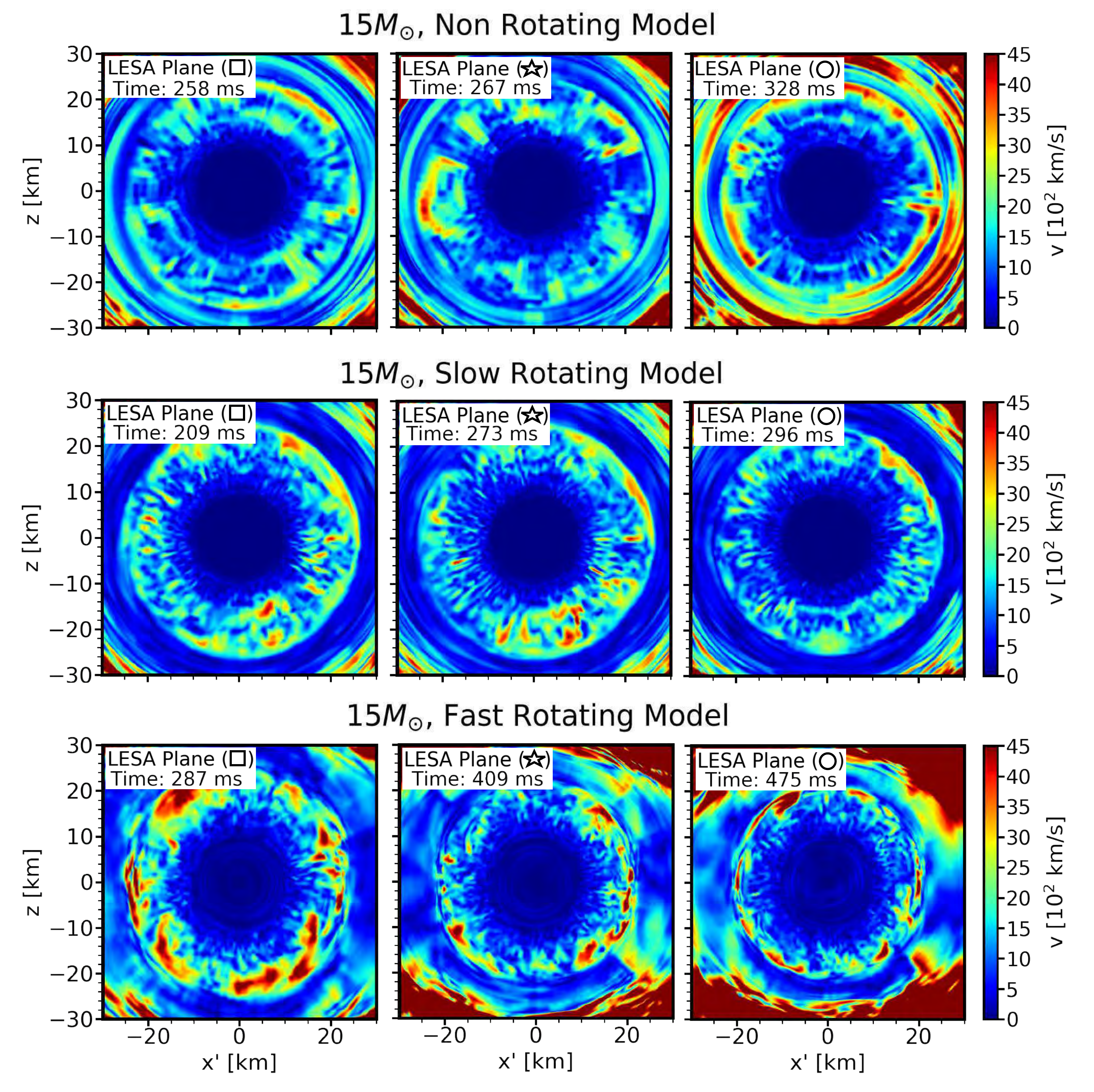}
\caption{Slices showing the spatial distribution of the absolute turbulent velocity of the fluid inside the PNS, as defined in Eq.~\ref{eq:abs_V},  at the times of the snapshots in Fig.~\ref{fig:LESA_sanpshots} for the non-rotating, slowly rotating, and fast rotating models from top to bottom, respectively.  Each cut is taken along the plane containing the ELN/LESA dipole and the rotational axis (see Fig.~\ref{fig:ELNaxis}). The plots are oriented such that the north and south poles are at the top and bottom, respectively. The convective activity is  confined to a narrower radial layer  as the time advances, reflecting the PNS contraction. This is especially visible for the rotating models.}
\label{fig:Vevo}
\end{figure*}

In order to test the above conjecture  in the presence of SN rotation, Fig.~\ref{fig:Vevo} displays the spatial pattern of the fluid velocity  at the times of the snapshots in Fig.~\ref{fig:LESA_sanpshots} for the non-rotating, slowly rotating, and fast rotating models from top to bottom, respectively. The spatial distribution is shown in the $(x^\prime,z)$ plane defined by the ELN/LESA dipole axis and the SN rotation axis (see Fig.~\ref{fig:ELNaxis}). In order to disentangle the convective activity from progenitor inherited or SASI-induced rotation of the PNS as well as contraction motion of the PNS, we subtract coherent velocity components from the total fluid velocity. The absolute turbulent velocity is then defined as 
\begin{equation}\label{eq:abs_V}
\text{v} = \sqrt{(v_r- \langle v_r\rangle)^2 + (v_\theta- \langle v_\theta\rangle)^2 + (v_\phi - \langle v_\phi \rangle)^2\ ,}
\end{equation}
where $v_{r, \theta, \phi}$ are the velocity components along the radial, latitudinal, and azimuthal basis vectors respectively, computed at each radius and angular zone of the simulation grid. The  volume averaged velocity components along $r, \theta,\phi$ ($\langle v_{r, \theta, \phi} \rangle$) are  defined as,
\begin{eqnarray}
\langle v_r\rangle &=& \frac{\sum_{\theta,\phi} v_r r^2 \sin{\theta} \Delta r \Delta \theta \Delta \phi }{\sfrac{4}{3} \;  \pi \; (r_3^3 - r_1^3)} \mathrm{,}\nonumber \\ \nonumber \\ \nonumber
\langle v_\theta \rangle &=& \frac{\sum_{\theta} v_\theta r^2 \sin{\theta} \Delta r \Delta \theta \Delta \phi }{r^2 \sin{\theta} \Delta r \Delta \theta \Delta \phi} \mathrm{,}\nonumber \\ \nonumber \\ 
\langle v_\phi \rangle &=& \frac{\sum_{\phi} v_\phi r^2 \sin{\theta} \Delta r \Delta \theta \Delta \phi }{2 \pi r^2 \sin{\theta} \Delta r \Delta \theta} \mathrm{,} 
\label{eq:avg_v_comp}
\end{eqnarray}
with $\Delta r, \Delta\theta, \Delta\phi$ the spacing of the simulation grid with mid-values $r, \theta$ and $\phi$ respectively, and $(r_3 - r_1) = \Delta r$, $r_{1,3}$ being the upper and lower radii defining the bin centered on $r$.

From Fig.~\ref{fig:Vevo}, one can see that the convective PNS layer is constrained to a narrower shell as time increases, reflecting  the PNS contraction.  However, the absolute turbulent velocity does not show any clear sign of an eventual weakening of the convective activity as the rotational velocity increases. Therefore, it cannot explain the suppression of the LESA dipole in the fast rotating model. 
It should be also noted that the LESA direction cannot  be always clearly identified in the snapshots of the convective velocities at a single time, given in Fig.~\ref{fig:Vevo}, because stochastic convective fluctuations can distort the time-averaged picture. Moreover, it is very difficult to subtract rotation effects in a clear way, in particular because massive polar downflows can stir the PNS activity in the fast-rotating model. A conclusive comparison among the three models is also hampered by the fact that the non-rotating model was computed with an angular resolution of  $4$ degrees  in contrast to the resolution of $2$ degrees adopted in  the two rotating models. Correspondingly, higher numerical viscosity is likely to damp the convective activity in the non-rotating model.   

Convection is driven along the gradient of the gravitational or effective (gravitational and centrifugal for the rotating models) potential, which is dominant in the radial direction. Hence, in order to quantify the effect of rotation on the  strength of the LESA instability, we explore the temporal evolution of the  turbulent kinetic energy in the PNS convection layer by focusing on its radial component. The radial component of the  turbulent kinetic energy in the PNS convection layer  is defined as
\begin{equation}\label{eq:rad_KE}
\mathrm{\mathcal{E}}_k(t) = \frac{1}{2}  \sum_{\theta,\phi}  \sum_{r_{\mathrm{min}}}^{r_\mathrm{max}}  \rho \mathrm{v}_r ^2 r^2 \sin\theta \Delta \theta \Delta \phi \Delta r \ ,
\end{equation}
with $\mathrm{v}_r  = (v_r- \langle v_r\rangle)$ and $\langle v_r\rangle$  was introduced in Eq.~\ref{eq:avg_v_comp}. For each time $t$, the radial component of the turbulent kinetic energy is computed in each angular zone and then  integrated  over the width of the PNS convection layer. The inner bound of the latter ($r_{\mathrm{min}}(\theta,\phi)$) is defined by the radius at which the absolute turbulent velocity (Eq.~\ref{eq:abs_V}) exceeds 500~km~s$^{-1}$. The outer bound of the PNS convection layer ($r_{\mathrm{max}}(\theta,\phi)$) is the radius at which the velocity steeply falls, reaching a local minimum  between  $\sim 20-25$~km  (see Fig.~\ref{fig:Vevo}). 

\begin{figure*}
\includegraphics[width=1.2\columnwidth]{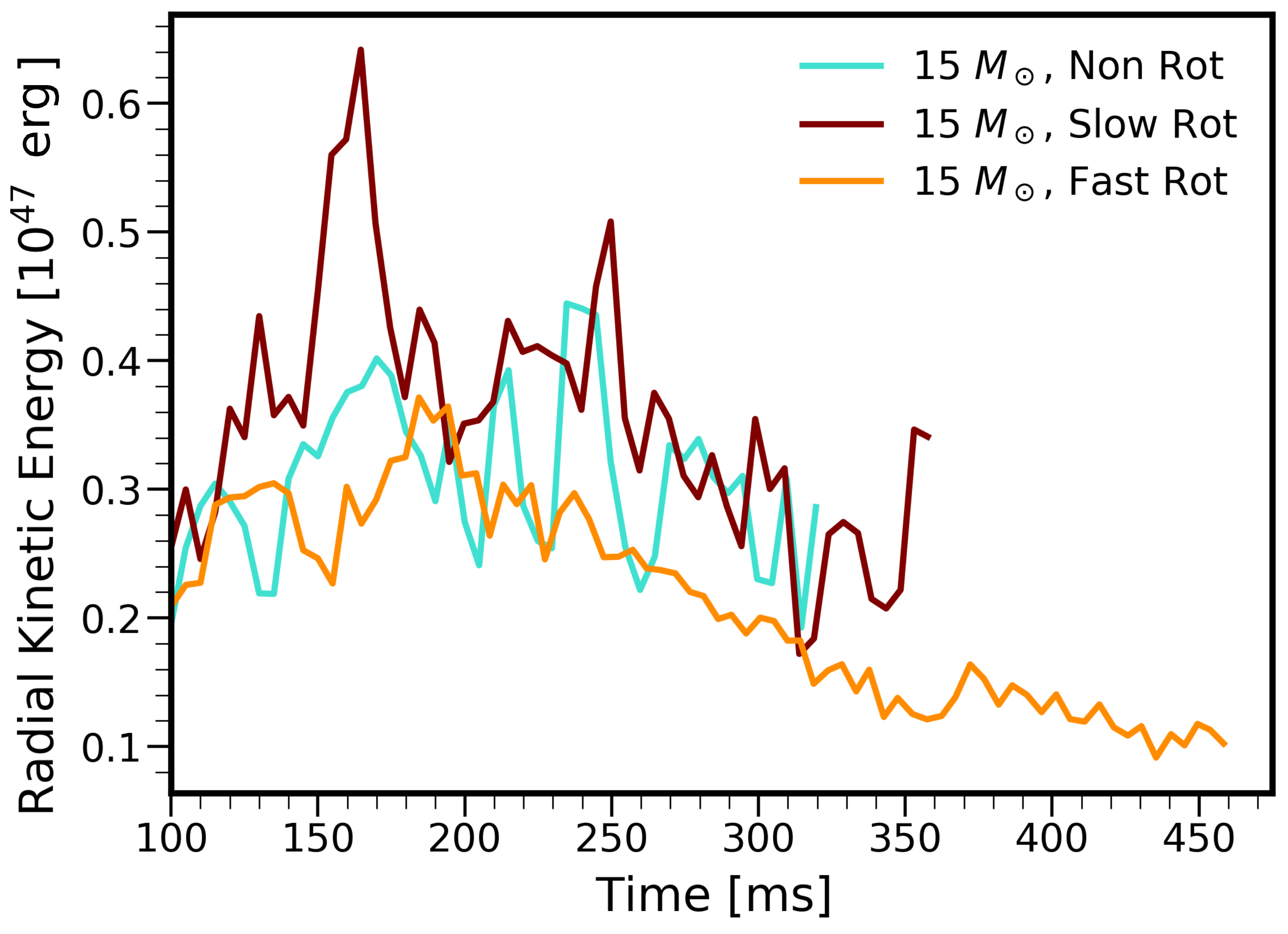}
\caption{ Radial component of the  turbulent kinetic energy in the PNS convection layer  as a function of time for the three 15 M$_\odot$ SN models (see Eq.~\ref{eq:rad_KE}). The radial component of the  turbulent kinetic energy  has been averaged over $5$~ms bins to smooth short-timescale fluctuations and  highlight the overall trend. Only post-bounce times later than $100$~ms, where the LESA dipole reaches a significant relative strength (see Fig.~\ref{fig:ELNdipole}), have been plotted.  The radial component of the kinetic energy in the PNS convection layer is reduced in the fast rotating model, implying that the amplitude of the LESA dipole may be linked to the strength of the radial convective mass motions. The clearly decreasing trend in the kinetic energy with time  reflects the contraction of the PNS convection layer as time progresses.}
\label{fig:KE}
\end{figure*}

The radial turbulent kinetic energy in the PNS convection layer, $\mathrm{\mathcal{E}}_k$, is shown in Fig.~\ref{fig:KE}  as a function of time for the three SN models for $t > 100$~ms post bounce (i.e., when the LESA dipole reaches a significant relative strength, see Fig.~\ref{fig:ELNdipole}). A running average over $5$~ms bins is used to smooth high-frequency fluctuations and  highlight the overall trend.   One can see that the rapid rotation is responsible for inducing a reduction of $\mathrm{\mathcal{E}}_k$. This, in turn, hints towards a connection between the strength of the LESA dipole  and the strength of the radial convective mass motions.  Note also that the radial turbulent kinetic energy decreases with  time for all models as a result of the PNS contraction.

\begin{figure*}
\includegraphics[width=2.\columnwidth]{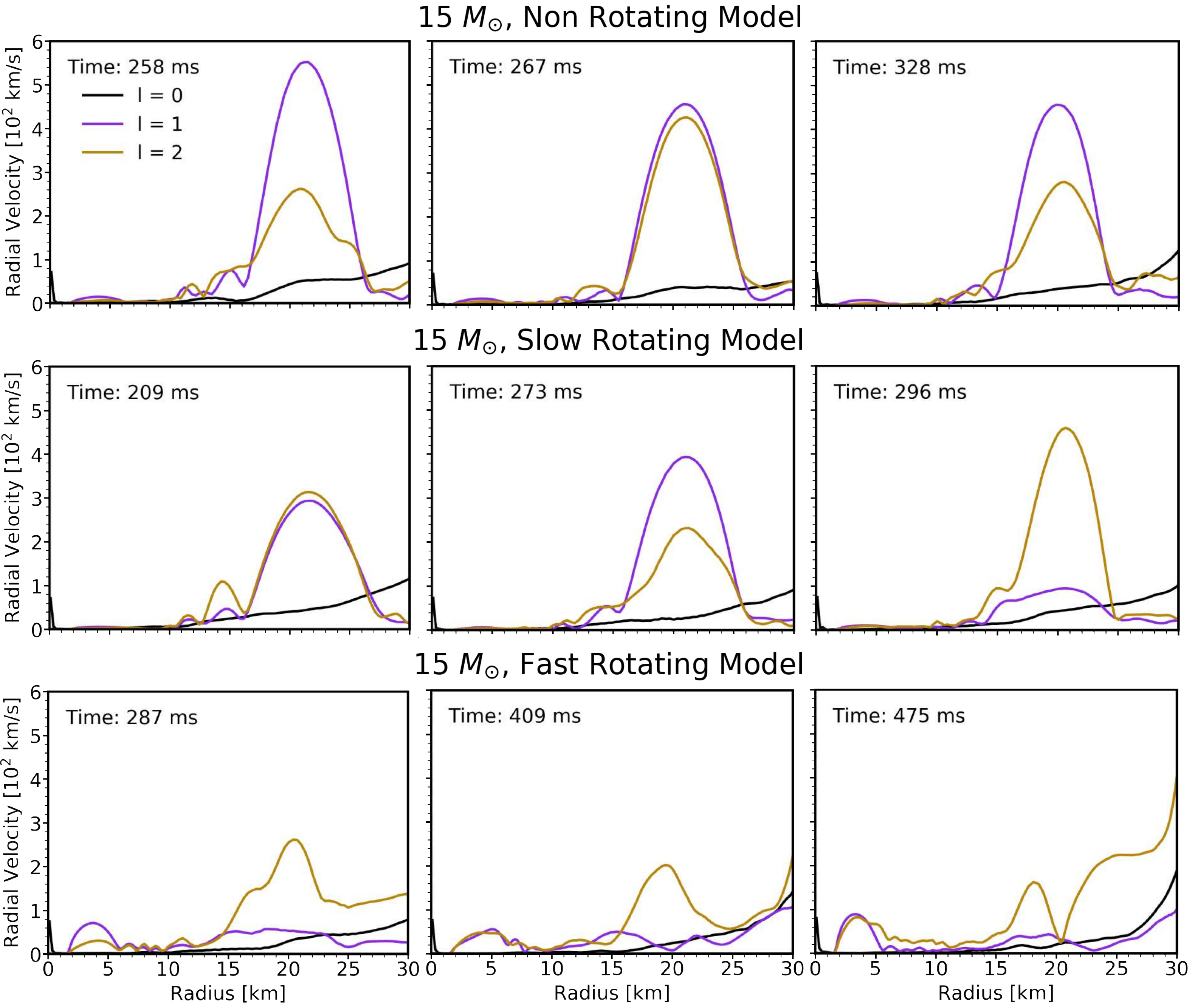}
\caption{Monopole, dipole, and quadrupole amplitudes of the radial component of the velocity, $v_r$, inside  the PNS for the non-rotating, slowly rotating, and fast rotating 15 M$_\odot$ SN models from top to bottom respectively, for the three time snapshots given in Fig.~\ref{fig:LESA_sanpshots}. Intervals  with a large dipole moment of $v_r$ in the PNS convection layer correlate with dominant LESA dipoles (see Fig.~\ref{fig:ELNdipole}), whereas intervals with a strong or dominant quadrupole of $v_r$ exhibit large quadrupoles of the ELN emission. With increasing
degree of rotation, the strength of convection is damped:
the multipole peak of the radial velocity field gets
lower, the width of the convective layer tends to shrink (the velocity
peak becomes more narrow), and a dominant dipolar asymmetry in
the case without rotation is replaced by a dominant quadrupolar
mode when rotation gets fast. All this is an unambiguous sign
that the LESA phenomenon (defined as a dipolar ELN emission asymmetry) 
is disfavored in the rapidly rotating model because of the strong
suppression of the $v_r$ dipole at all radii within the PNS.}
\label{fig:V_dipoles}
\end{figure*}

The strength of LESA in the (non-)rotating progenitors can be inferred by investigating the dipole and quadrupole moments of the radial velocity component, $v_r$, within the PNS (see Refs.~\cite{Glas:2018vcs,Powell:2018isq}). Figure~\ref{fig:V_dipoles} shows the amplitudes of the monopole, dipole, and quadrupole moments of $v_r$ for the three 15 M$_\odot$ SN models as a function of radius at each time snapshot given in Fig.~\ref{fig:LESA_sanpshots}. 
The monopole (plotted in black) corresponds to the slow overall contraction of the PNS, the higher multipoles are components of the mode pattern of the convective motions. These are superimposed in the fast rotating model by effects connected to the rapid rotation, which leads to a time-dependent centrifugal deformation and thus non-vanishing radial velocity components in the convectively stable PNS core and near the PNS surface.

For radii smaller than $\sim 25$~km, the $v_r$ dipole and the quadrupole are always larger than the monopole. 
The non-rotating model shows a leading $l=1$ mode. Instead, in the slow rotating model, the $l=2$ mode grows between the first and the last considered time snapshots, because rotation has a larger effect at later times. Simultaneously,  the $l=1$ mode decreases in amplitude, although some of the visible relative variations of the dipole and quadrupole may be caused by stochastic fluctuations of the convective activity. In the fast rotating model, the dipole and the quadrupole have comparable amplitudes at small radii ($r \lesssim 10$~km), but the $l=2$ mode clearly dominates at larger radii for all the three snapshots. 
The non-vanishing low-amplitude activity interior to the convective shell in this model is a consequence of the rotational deformation of the PNS (causing quadrupolar contributions), and of PNS core motion in response to the impact of massive polar accretion downflows (adding a dipolar component). 

Comparing the three SN models, one can notice that, as the rotational speed increases, the strength of the dipole of the radial  velocity in the PNS  decreases, disfavoring the development of the LESA phenomenon. Consequently, dipolar anti-correlation between the $\nu_e$ and $\bar{\nu}_e$ luminosities does not arise and the formation of long-lasting hemispheric asymmetries inside the PNS is disfavored.  These effects were previously illustrated in Figs.~\ref{fig:relatives}, \ref{fig:ELNdipole} and \ref{fig:Yeevo} and confirm that the progenitor rotation inhibits the growth of LESA.

The decrease of the radial component of the turbulent kinetic energy and  the reduced strength of the dipole of the radial  velocity in the PNS are clearly visible in Figs.~\ref{fig:KE}, \ref{fig:V_dipoles}  as the rotational velocity increases. We stress that  the correlation of this effect of rotation with the reduced LESA amplitude in the rotating models is currently a purely empirical finding. However, the interpretation of LESA as a consequence of the preferred growth of low spherical harmonics modes of the flow in the PNS convective layer  would offer a natural explanation (see also Ref.~\cite{Glas:2018vcs}). In fact, the Chandrasekar theory~\cite{1961hhs..book.....C} predicts a suppression of the $l=1$ mode for faster rotation. Since the Taylor number in the PNS convection zone of our fast rotating model is $\mathcal{O}(10^8)$, this should naturally lead to a suppression of the growth of the dipole flow.

The weaker growth and lower amplitude of the LESA dipole for faster rotation in this context require further investigation by future work. This is beyond the scope of the present paper, which  primarily aims to identify the imprints of rotation in the neutrino signal.

\section{Conclusions}\label{sec:conclusions}

Throughout this work we investigated the effects of progenitor rotation on the development of neutrino-emission asymmetries associated with the standing accretion shock instability (SASI) and the lepton-emission self-sustained asymmetry (LESA)  in core-collapse SNe.  Given the non-trivial interplay of the hydrodynamical instabilities, our work is a first attempt to classify the effects of rotation on the development of SASI and LESA through their imprints in the neutrino signal. This study was carried out with three self-consistent models of a 15 M$_\odot$ SN; one non-rotating model and two models with rotational velocities within observational constraints. 

The non-rotating model is characterized by a SASI phase imposing large-amplitude sinusoidal modulations on the neutrino emission properties.
As a consequence of the SN rotation, the slowly rotating model develops  spiral SASI activity that is much weaker than in the non-rotating model, and the neutrino emission properties show  modulations with smaller amplitude than in the non-rotating model. The rapidly rotating model, instead, has a  strong spiral SASI that aids the stellar explosion. Nevertheless, in this last case, the early SASI activity leaves only rather weak imprints on the neutrino signal.   The neutrino signal of the  fast rotating model, instead, features prominent signatures of the long-lasting polar downflows that develop in the post-explosion phase due to the rapid rotation, and temporal variations hardly depend on the angular location of the observer (see Figs.~\ref{fig:propertiesFR}, \ref{fig:relatives}).

Our findings suggest that the SN progenitor rotation also disfavors the growth of the LESA instability by weakening the convective activity within the PNS along the radial direction. 
In fact, the radial component of the turbulent kinetic energy in the PNS convection layer is lower when the rotational velocity increases, and  the strength of the dipole of the radial component of the fluid velocity  inside the PNS decreases. As a consequence, the  development of LESA as a dipolar ELN asymmetry is inhibited.  This is especially evident in the fast rotating model where we do not observe any sizeable anti-correlation between the $\nu_e$ and $\bar{\nu}_e$ luminosities (see Fig.~\ref{fig:relatives}) nor the LESA-characteristic hemispheric asymmetry of the electron fraction in the deleptonized layer around the PNS convection zone (see Fig.~\ref{fig:Yeevo}). We conclude that the transiently large dipole (between $\sim 100$~ms and $\sim 200$~ms, post-bounce) of the lepton number emission (see Fig.~\ref{fig:ELNdipole}) in the fast rotating model is a consequence of the strong SASI spiral motions and the corresponding asymmetric PNS accretion. At later stages, after the onset of the explosion in this model, quadrupolar asymmetry dominates the lepton number emission, as well as the luminosities of $\nu_e$ and $\bar{\nu}_e$ (see Fig.~\ref{LESA_SASI_Mulitpoles}). This can be understood by the accretion luminosity associated with the two polar downflows. 

Our results provide empirical evidence that the growth of LESA is interestingly inhibited by progenitor rotation. A quantitative theory of  the suppression of  convective activity along the radial direction in the PNS and its impact on the development of the LESA phenomenon requires further work.      
Our findings also shows the power intrinsic to the neutrino signal in providing hints on the SN core dynamics. In fact, together with gravitational waves, neutrinos carry imprints of the physics affecting the SN mechanism in the innermost regions not accessible otherwise.

\section*{Acknowledgments}
We are grateful to Tobias Melson for his assistance in evaluating the data of Ref.~\cite{Summa:2017wxq}. This project was supported by the Villum Foundation (Project No.~13164), the Knud H\o jgaard Foundation, the European Research Council through grant ERC-AdG No.\
341157-COCO2CASA, the Deutsche Forschungsgemeinschaft through Sonderforschungbereich
SFB~1258 ``Neutrinos and Dark Matter in Astro- and
Particle Physics'' (NDM), and the Excellence Cluster ``Universe''
(EXC~153). The model calculations were performed on SuperMUC at
the Leibniz Supercomputing Centre with resources granted
by the Gauss Centre for Supercomputing (LRZ project ID: pr74de).

\appendix
\section{Direction dependent variations of the neutrino emission properties}\label{sec:appendix}

In Sec.~\ref{sec:SASIconv} (Eq.~\ref{eq:deltarho}), the $\Delta$ and $\rho$ parameters are introduced to disentangle the variations of the neutrino properties due to SASI from the ones due to LESA, i.e.~ the modulations of the neutrino signal and its spread in intensity between the different angular zones. To that purpose,  the time-dependent  ``mid-line luminosity'' in each angular direction ($\tilde{L}_{\bar\nu{_e}}(t,\theta,\phi)$) is  used to compute the mean absolute  modulation amplitude  of the time-dependent $\bar\nu_e$ luminosity as well as the mean absolute spread in luminosity over a particular time interval. 

In this Appendix, we outline the interpolation method employed to find $\tilde{L}_{\bar\nu{_e}}(t,\theta,\phi)$, and illustrate the difference in the $\Delta$ and $\rho$ parameters. We use  the 27 and 11.2 M$_\odot$ models as examples because they have the most extreme variations of the $\Delta$ ($27\,M_\odot$) and $\rho$ ($11.2\,M_\odot$) parameters, see Sec.~\ref{sec:SASIconv}. For the sake of simplicity, we will focus on $L_{\bar\nu{_e}}$; the luminosity of any other neutrino flavor shows  similar trends.

\begin{figure*}
\includegraphics[width=1.6\columnwidth]{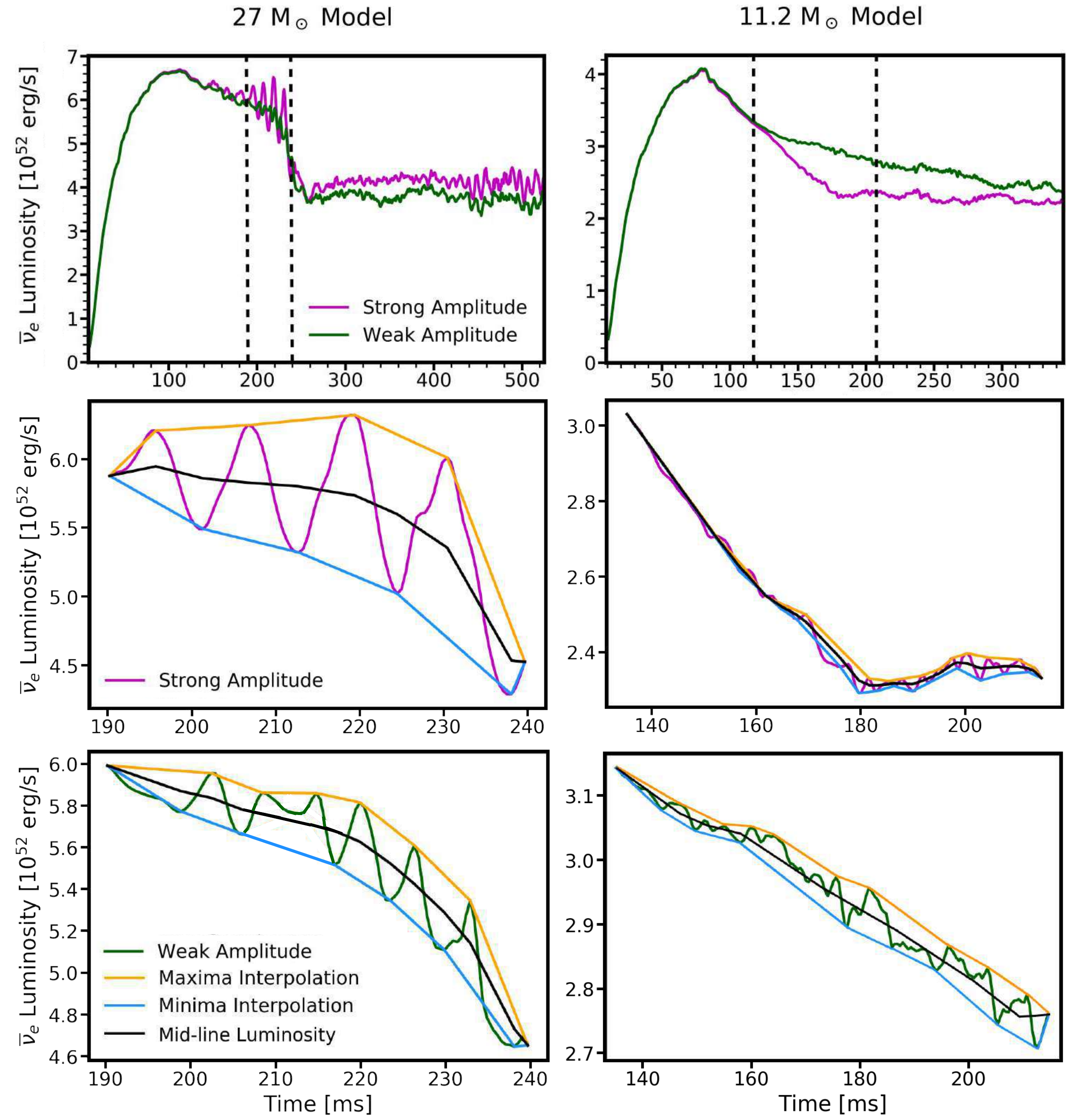}
\caption{Representative examples of the method used to find the time-dependent mid-line luminosity $\tilde{L}_{\bar\nu{_e}}(\theta,\phi)$ for each SN model. The $27$ and $11.2\,M_\odot$ models are shown in the panels on the  left and right, respectively. Two observer directions corresponding to strong (in magenta) and weak (in green) modulations of  $L_{\bar\nu{_e}}$ have been selected. The time interval selected for the interpolation is indicated by dashed black lines (top panels). The mid-line luminosity obtained is given by the solid black line for the strong (middle panels) and weak (bottom panels) modulation directions.}
\label{fig:Midline_int}
\end{figure*}

We first select a time window, $[t_1, t_2]$, for which the  modulation amplitude of $L_{\bar\nu{_e}}$ is on average the strongest along all angular directions. Then, we determine the local maxima and minima of $L_{\bar\nu{_e}}$  within this time window. For each angular direction, the  time-dependent mid-line luminosity $\tilde{L}_{\bar\nu{_e}}(\theta,\phi)$ is found by first interpolating among all the local maxima and all the local minima, and then by taking the mid value between the interpolated curves of the minima and maxima at each time step. Figure~\ref{fig:Midline_int} illustrates this  method for the two SN models considered in this Appendix. The top panels show the luminosities along the directions of weakest and strongest modulation amplitudes of the neutrino signal. The time window considered is marked with dotted black lines. The bottom two panels show the mid-line interpolation of each direction for each model within $[t_1, t_2]$. 

 The $\Delta$ parameter is computed for all angular directions by taking the average distance between $L_{\bar\nu{_e}}$  and $\tilde{L}_{\bar\nu{_e}}$ over all time steps in $[t_1,t_2]$. Thus, $\Delta$ gives a measure of the mean total  modulation  amplitude of the neutrino luminosity due to SASI. 
 
 The $\rho$ parameter is instead computed by first finding the 4$\pi$-average of the mid-line luminosities, $\langle \tilde{L}_{\bar\nu{_e}}\rangle$. Then, we estimate the mean deviation of $\tilde{L}_{\bar\nu{_e}}(\theta,\phi)$ from $\langle \tilde{L}_{\bar\nu{_e}}\rangle$ in $[t_1,t_2]$. Thus, $\rho$  gives an indication of the overall spread in the average neutrino luminosity $L_{\bar{\nu}_e}$ over all directions  due to LESA. 

\begin{figure*}
\includegraphics[width=1.6\columnwidth]{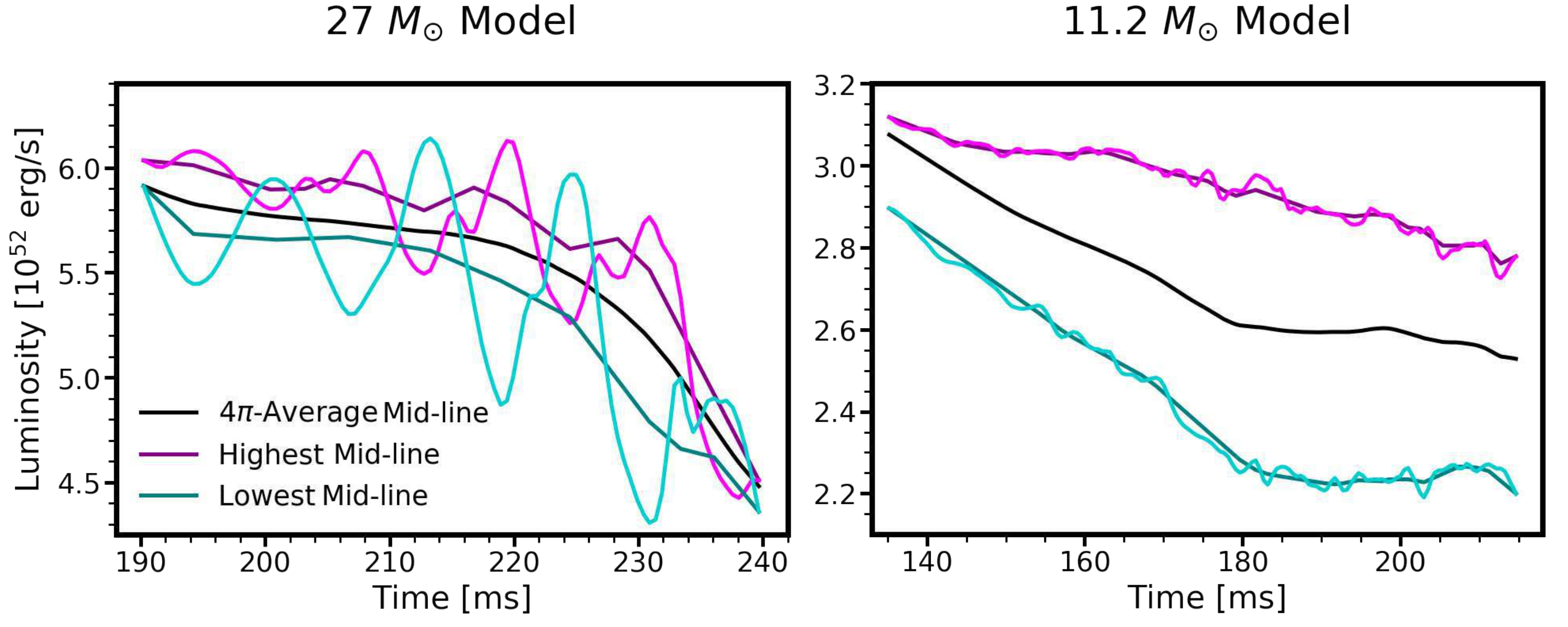}
\caption{Highest average (teal), lowest average (magenta), and 4$\pi$-average (black) mid-line luminosity signals for the $27$ and $11.2\,M_\odot$ models from left to right respectively. Each highest and lowest mid-line luminosity is embedded in the corresponding luminosity signal. The $27\,M_\odot$ model shows large amplitude ($\Delta$) modulations, while the $11.2\,M_\odot$ model shows a wide spread in the overall neutrino luminosity between different angular directions ($\rho$).}
\label{fig:RHO}
\end{figure*}

Figure \ref{fig:RHO} shows $L_{\bar\nu{_e}}$ and the corresponding mid-line luminosities  $\tilde{L}_{\bar\nu{_e}}(\theta,\phi)$ with the overall highest (teal) and lowest (magenta) values over angular directions  for both SN models, along with the 4$\pi$-average mid-line luminosity $\langle \tilde{L}_{\bar\nu{_e}}\rangle$. From this Figure, it becomes clear that in the $11.2\,M_\odot$ model, the distance from the mid-line luminosity to its 4$\pi$-average for a particular angular direction  is much larger than the average  modulation  amplitude of the neutrino luminosity for fixed angular direction, meaning that this model will exhibit strong $\rho$ modulations and relatively low $\Delta$ modulations. On the contrary, in the $27\,M_\odot$ model, the magnitude of the amplitude modulations for fixed angular direction dominates the spread in mid-line luminosity, such that $\Delta$ modulations are dominant.

It should be noted that the interpolation method involves certain approximations. For instance, differentiating local minima and maxima in the neutrino luminosity due to legitimate physical  changes from small-scale noise fluctuations is essential. Because the same maxima/minima parameters are applied to all angular directions, but the size of small-scale fluctuations due to noise may differ between angular directions, there may be instances where the interpolation misses or includes an undesired local minimum or maximum.  To minimize the error coming from this effect, we take the integrated average over all times within $[t_1,t_2]$ and apply this at all angular directions. We also stress that $\tilde{L}_{\bar\nu{_e}}$ and $\langle \tilde{L}_{\bar\nu{_e}}\rangle$ are fictitious quantities. They have only been introduced to characterize the spread in the neutrino luminosity between the different angular directions. 


\bibliography{SNrotating}

\end{document}